\documentclass[a4paper,12pt]{article}

\usepackage{multicol}

\usepackage{arydshln}
\usepackage{graphicx}
\usepackage{subfigure}
\usepackage{here}
\usepackage{float}
\usepackage{amsthm}
\usepackage{amssymb,amsmath}
  \usepackage{natbib}
 \usepackage{MnSymbol}

\usepackage[utf8]{inputenc}
\usepackage[T1]{fontenc}
\usepackage[frenchb,english]{babel}
\usepackage{mathrsfs}
\usepackage{enumerate}
\usepackage{color}
\usepackage[dvipsnames]{xcolor}
\usepackage{colortbl}

\usepackage[colorlinks=false,linkcolor=blue]{hyperref}
\usepackage[nottoc, notlof, notlot]{tocbibind}
\usepackage{titling}
\usepackage{tipa}
\usepackage[sc]{caption}
\usepackage[]{pdfpages}
\newcommand{\subtitle}[1]{%
  \posttitle{%
    \par\end{center}
    \begin{center}\large#1\end{center}
    \vskip0.5em}%
}
\usepackage{geometry}
\usepackage{bm}

\usepackage{pgf,tikz}
\usepackage{mathrsfs}
\usetikzlibrary{arrows}
\usetikzlibrary{decorations.pathreplacing}

\newcommand{\edgexy}{{\langle x,y \rangle}}

\renewcommand{\epsilon}{\varepsilon}

\usepackage[plain,noend,boxed]{algorithm2e}
\usepackage{listings}

\theoremstyle{definition}

\usepackage{chngcntr}

\title{A Phylogenetic Model of the Evolution of Discrete Matrices for the Joint Inference of Lexical and Phonological Language Histories}

\author{Gr\'{e}goire Clart\'{e}, Robin J. Ryder}

\begin{document}

\maketitle

\begin{abstract}
We propose a model of the evolution of a matrix along a phylogenetic tree, in which transformations affect either entire rows or columns of the matrix. This represents the change of both lexical and phonological aspects of linguistic data, by allowing for new words to appear and for systematic phonological changes to affect the entire vocabulary. We implement a Sequential Monte Carlo method to sample from the posterior distribution, and infer jointly the phylogeny, model parameters, and latent variables representing cognate births and phonological transformations. We successfully apply this method to synthetic and real data of moderate size.
\end{abstract}


\section{Introduction}


Languages change and diverge through time in a way that has often been compared to the evolution of biological species. The history of related languages, at a depth of a few millennia, is often represented by a dated phylogenetic tree. These trees have traditionally been reconstructed by hand by expert linguists, using the comparative method \cite{campbell2013historical} to disentangle lexical, phonological and morphosyntactic evidence and deduce which languages are close cousins.
In their pioneering work, \citet{GrayRussellAtkinsonQuentin} introduced quantitative methods to infer the phylogeny of the Indo-European languages, using a model built from genetics. The data they used are a binarized form of lexical data, recording presence or absence of \emph{cognate classes} (defined below). This work has since led to the creation of many models specifically tailored to cognate data \citep{Nicholls2007dated,chang2015ancestry,Bouckaert2012Mapping,rama2016ancestry}, which have been applied to Dravidian \citep{Kolipakam18Dravidian}, Sino-Tibetan \citep{Sagart10317}, Pama-Nyungan \citep{bowern2012computational}, or Austronesian languages \citep{Gray2009Austronesian}, amongst others. We refer the reader to \citet{hoffmann2021bayesian} for a review.


Most phylogenetic methods used in computational linguistics thus rely on lexical datasets. The raw data are word forms attested in each language for a number of pre-specified meanings. These raw data are then transformed into a binary form, called cognacy classes: a cognacy class corresponds to a set of words in different languages with provably shared history. For example, the English \emph{father}, German \emph{Vater} and Latin \emph{pater} are cognates. In binary format, the column corresponding to this cognacy class would bear a $1$ for these three languages.
The similarity between these, and other, words, is evidence of shared history between these languages.

The transformation of raw word forms into cognacy classes is a time-consuming process for expert linguists of the language family under study. It typically transforms the raw data into a binary matrix recording the presence or absence of each cognacy class in each language; the binary matrix is the only aspect of the data used in the statistical inference. This allows for easier stochastic modelling, but at the same time it loses an important part of the information contained in the raw data, since the phonological aspects of the data are lost. 

Models of the evolution of other aspects of languages, including phonology, remain scarce, even though this aspect of the data is also highly informative of language history. \cite{List2017Potential} describe a method to infer cognates automatically, but this is done independently from the inference of the phylogeny; \cite{Bouchard2013Automatic} reconstruct phonological changes, but their method requires a previously reconstructed phylogeny as input. In a similar vein, Machine Learning algorithms have been applied to the task of cognate prediction, which proposes a plausible form for a word in a language based on related languages; see \citet{fourrier2022probing} and the recent survey by  \citet{dekker2020word}. These works do not attempt to jointly infer the phylogeny of the languages.

A key aspect of cognate-only data is that they are intrinsically one-dimensional: the observed data at a single language consist of a vector of presence/absence of each cognate. (The whole data set with several languages can then be seen as a matrix, with one language per row.) Most models then assume that the vector elements evolve independently, allowing for likelihood evaluations through a  pruning algorithm \cite{Felsenstein}. 

 This paper describes the first attempt at modelling jointly the changes in cognates and phonology, thus describing the evolution of the whole word form and not just the cognacy classes. In our setting however, the data at a single language are two-dimensional: they consist of a matrix, with each row corresponding to a meaning, and each column corresponding to a phoneme. The set of phonemes is finite and known; in this paper, we assume that the data are aligned, in a way discussed later on. The whole data set  with several languages is thus a 3-dimensional array, and each array entry is a phoneme.
 Indeed, the lexicon of a language is modified from its parent language via two major processes: (1) a new word may arise, corresponding to the birth of a new cognacy class; (2) existing words may be modified by a phonological change, such as $/p/\to /f/$. Process (1) is well modelled by existing works; for matrix data, it corresponds to the replacement of all values in a row of the matrix. The difficulty with process (2) is that it typically affects simultaneously many meanings in the language \citep{campbell2013historical}: several values in a column of the matrix will be modified.

By modelling both the arisal of new cognates and the phonological changes, we are able to infer jointly the language phylogeny, cognate births and deaths, and phonological changes. We present a stochastic model of the evolution of a discrete matrix on a tree in Section \ref{sec:Model}, and describe in Section \ref{sec:Numericalmethods} numerical methods to sample from the associated posterior. The application motivating this work is the reconstruction of the history of Polynesian languages, which we describe and analyse in Section \ref{sec:appli}, along with examples on simulated datasets.

\section{Background},

\colorlet{col1}{OliveGreen}
\colorlet{col2}{Orange}
\colorlet{col3}{RoyalPurple}
\colorlet{col4}{Red}
\colorlet{col5}{TealBlue}
\colorlet{col6}{Gray}
\colorlet{col7}{NavyBlue}
\colorlet{col8}{Rhodamine}
\colorlet{col9}{BrickRed}

\begin{table}[t]
\centering
\begin{tabular}{|c|c|c|c|c|c|c|}
\hline
meaning & \itshape eye &\itshape seven &\itshape hand &\itshape flower &\itshape to breathe &\itshape fruit\\\hline\hline
Samoan & \color{col1}{mata} & \color{col2}{fitu} & \color{col3}{lima} & \color{col4}{fuŋa} & \color{col6}{mānava} & \color{col9}{fua}\\\hline
Tahitian & \color{col1}{mata} & \color{col2}{hitu} & \color{col3}{rima} & \color{col5}{pua} & \color{col7}{huti} & \color{col9}{hua}\\\hline
Hawaiian & \color{col1}{maka} & \color{col2}{hiku} & \color{col3}{rima} & \color{col5}{pua} & \color{col8}{hanu} & \color{col9}{hua}\\\hline
\end{tabular}
\caption{Toy lexical dataset $D$ for three Austronesian languages: Samoan, Tahitian and Hawaiian. Cognates are displayed in the same colour. Data adapted from \cite{abvd}.\label{tab:lexdata}}
\end{table}

\begin{table}[t]
\centering
\begin{tabular}{|c|>{$\{$}c<{$\}$}|}
\hline
eye$_1$ & \color{col1}{mata, mata, maka}\\\hline
seven$_1$ & \color{col2}{fitu, hitu, hiku} \\\hline
hand$_1$ & \color{col3}{lima, rima, rima}\\\hline
flower$_1$ & \color{col4}{fuŋa}\\
flower$_2$ & \color{col5}{pua, pua}\\\hline
breathe$_1$ & \color{col6}{mānava}\\
breathe$_2$ & \color{col7}{huti}\\
breathe$_3$ & \color{col8}{hanu}\\\hline
fruit$_1$ & \color{col9}{fua, hua, hua}\\\hline
\end{tabular}
\caption{Cognate sets $D / \hcrossing$ for the data of Table \ref{tab:lexdata}.\label{tab:cogsets}}
\end{table}

\newcommand{\myrotate}[1]{\rotatebox{90}{#1}}
\begin{table}[t]
\centering
\begin{tabular}{|c||>{\color{col1}}c||>{\color{col2}}c||>{\color{col3}}c||>{\color{col4}}c|>{\color{col5}}c||>{\color{col6}}c|>{\color{col7}}c|>{\color{col8}}c||>{\color{col9}}c||}
\hline
 & \myrotate{eye$_1$} & \myrotate{seven$_1$} & \myrotate{hand$_1$} & \myrotate{flower$_1$} & \myrotate{flower$_2$} & \myrotate{breathe$_1$} & \myrotate{breathe$_2$} & \myrotate{breathe$_3$} & \myrotate{fruit$_1$}\\\hline
 Samoan & 1&1&1&1&0&1&0&0&1\\
 Tahitian & 1&1&1&0&1&0&1&0&1\\
 Hawaiian & 1&1&1&0&1&0&0&1&1\\\hline
\end{tabular}
\caption{Transformed data $\tilde D$, based on the data of Table \ref{tab:lexdata} listing presence/absence of each cognacy class.\label{tab:cogmatrix}}
\end{table}

The vast majority of quantitative studies for Historical Linguistics rely on cognate data, which are a transformation of lexical data. A set of $L$ languages  and of $M$ meanings are chosen in advance. Linguists collect the word form(s)
 for each language $\ell$ and meaning $m$. This gives rise to raw data $D$, in the form of a $L\times M$ matrix of words, exemplified in Table \ref{tab:lexdata}. We say two words $w=D_{\ell,m}$ and $w'=D_{\ell',m}$ in two different languages are cognates, and write $w\hcrossing w'$, if they have the same meaning and have evolved from a common ancestral form. For example, the Samoan, Tahitian and Hawaiian words for \emph{eye} are cognates: "mata"$\hcrossing$"mata"$\hcrossing$"maka". Similarly, the three words for \emph{seven} are cognates: "fitu"$\hcrossing$"hitu"$\hcrossing$"hiku". This is shown by noticing not only the similarities between the words, but also that differences occur in a systematic fashion: an $/f/$ in Samoan often corresponds to an $/h/$ in Tahitian and Hawaiian (see \emph{fruit}), and a $/t/$ in Samoan and Tahitian often corresponds to a $/k/$ in Hawaiian (see \emph{eye}). One possible explanation would be that in the past, there was an $/h/\to/f/$ phonological transformation on the branch leading to Samoan, after it diverged from Tahitian and Hawaiian, and a $/t/\to/k/$ transformation on the branch leading to Hawaiian.
 
 Since $\hcrossing$ is an equivalence relation, we can define equivalence classes for $\hcrossing$, called cognacy classes, with $c_m$ classes for meaning $m$; this corresponds to the quotient space $D/\hcrossing$ and an example is given in Table \ref{tab:cogsets}. Column $m$ of $D$ is then transformed into a $L\times c_m$ binary matrix, where a 1 in column $i$ for language $\ell$ means that cognacy class $i$ is represented in language $\ell$, and a 0 means it is not represented in language $\ell$. We concatenate these binary matrices to obtain the transformed data $\tilde D$, a $L\times \sum_m c_m$ binary matrix, as shown in Table \ref{tab:cogmatrix}. It is $\tilde D$ which is then used in the statistical analysis. Note that in $\tilde D$, we typically do not use the original meanings, and in particular we ignore the fact that certain columns of $\tilde D$ correspond to the same meaning; we then assume that the columns of $\tilde D$ are iid.
 
 The transformation $D\mapsto \tilde D$ is a time-consuming process performed by hand by linguists expert of the language family under study. This binarization also discards much potentially useful information. In particular, the phonological transformations (which contain signal on the phylogeny) could be recovered from $D$, but not from $\tilde D$.
 
In the format $\tilde D$, the data observed at a leaf language are assumed to be a binary vector of cognate presence/absence, with vector elements assumed to be independent. The models can thus be represented by a generative process with state space $\{0, 1\}$, of which we observe multiple independent realizations. In our work however, the data at a single language are two-dimensional: they consist of a matrix, in which each row corresponds to a meaning, and each column corresponds to a phoneme. The set of phonemes is finite and known. The whole data set  with several languages is thus a 3-dimensional array, and each array entry is a phoneme.

Formalizing our notation, let $D_\ell$ be a matrix representing the data observed at 
 leaf language $\ell$. Table \ref{tab:phondata} shows how some of the raw data of Table \ref{tab:lexdata} are represented. The value of $D_\ell(m,k)$ corresponds to the $k$th phoneme of the word for meaning $m$. It belongs to a known discrete set  identified with the integers $\{1\ldots n_k\}$. The number of possible elements $n_k$ may vary between columns; this is the case in our application of Section \ref{sec:appli}.
 
 \begin{table}
\begin{tabular}{ccc}
\bf Samoan & \bf Tahitian & \bf Hawaiian\\
\begin{tabular}{|>{\itshape}c||c|c|c|c|}
\hline
eye & m&a&t&a\\\hline
seven & f&i&t&u\\\hline
hand & l&i&m&a\\\hline
\end{tabular}
&
\begin{tabular}{|>{\itshape}c||c|c|c|c|}
\hline
eye & m&a&t&a\\\hline
seven & h&i&t&u\\\hline
hand & r&i&m&a\\\hline
\end{tabular}
&
\begin{tabular}{|>{\itshape}c||c|c|c|c|}
\hline
eye & m&a&k&a\\\hline
seven & h&i&k&u\\\hline
hand & r&i&m&a\\\hline
\end{tabular}
\end{tabular}
\caption{Representation of part of the data of Table \ref{tab:lexdata} as a collection of aligned matrices.\label{tab:phondata}}
\end{table}

\section{Model}
\label{sec:Model}
We propose a generative model of the evolution of a discrete matrix  on a rooted and dated binary tree. 

\subsection{Data format}

We consider data collected at $L$ leaves $\ell=1\ldots L$. At leaf language $\ell$, the data are a $M\times K$ matrix, so that  $D_\ell(m,k)$ is the value for character $k$ of the meaning $m$ in language $\ell$. The set of possible values for $D_\ell(m,k)$ is assumed known and finite; we identify it with $1,\ldots,n_k$. Different characters may have different sets of possible values.

\subsection{Evolution model}

\label{ling:sec:modele}

We represent the vocabulary at point $x$ on the tree by a matrix $U_x$. At the root, the word forms are drawn randomly from the uniform distribution $\pi_0$ over all forms. At a branching event on the tree, the two child branches inherit the vocabulary of the parent branch, and then evolve independently. We model word change by two independent Poisson processes on a phylogenetic tree; these processes modify $U$. These processes are:

\begin{itemize}
\item a creation process: at a creation event, a new word is born. A meaning is drawn at random and the entire corresponding row is replaced, with new values drawn from the uniform distribution $\pi_0$.
Creation events arise from a Poisson Process with rate $\mu$, independently for each meaning. 
This process is similar to the $M_k$ model \citep{pagel1994detecting,lewis2001likelihood}.

\item a transformation process: at a transformation event, a phonological change $r$ takes place, of the form $r:/a/\to/a'/$, affecting phoneme position $k$. In column $k$, across all meanings, each occurrence of $/a/$ is replaced by $/a'/$ with probability $q_k$ independently.  For each character $k$, transformation events arise from a Poisson process with rate $\lambda_k$. The change $r$ is then drawn from a catogerical distribution over a predetermined set of possible transformations $\mathfrak{R}_k$. This is a  new adjunction to the language phylogeny literature, as we are aiming at simulating the natural evolution of phonemes through time, as described by phonological rules \citep{Hock1991Principles}. This model, to our knowledge is not similar to another one existing in the literature, because of the regularity of the transformations.

\end{itemize}

We will identify a transition $r$ with its transition matrix, defined for $a \rightarrow b$ as $r_{i,j} = \left\lbrace \begin{matrix}
1 & \text{if } i=j, i\neq a \\
\beta_k & \text{if } i=a, j=b \\
1-\beta_k & \text{if } i=a, j=a \\
0 & \text{otherwise}
\end{matrix} \right.$.

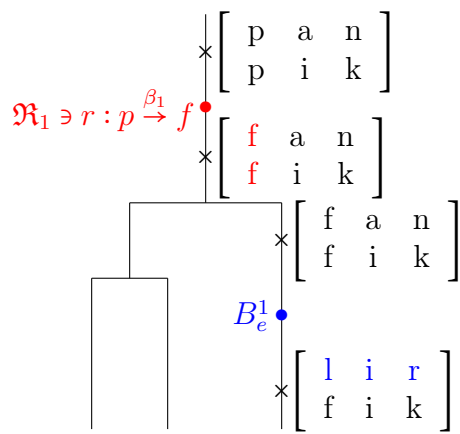
\begin{figure}
\center
\caption{Stochastic evolution processes. The first event corresponds to a phonological transformation $/p/\to /f/$: in column 1 of $U_x$, all occurrences of  $/p$ may transform into $/f/$. The second event corresponds to the creation of a new cognate: the second row of of $U_x$ is entirely renewed.}
\label{treeempr}
\begin{tikzpicture}[scale=0.5]
\draw (0,0) -- (0,5);
\draw (-2,0) -- (2,0);
\draw (-2,0) -- (-2,-2);
\draw (-3,-2) -- (-1,-2);
\draw (-3,-2) -- (-3,-6);
\draw (-1,-2) -- (-1,-6);
\draw (2,0) -- (2,-6);

\draw (0,4) node [anchor =west] {$\left[\begin{tabular}{ccc}
p & a & n \\ p & i & k
\end{tabular}\right]$};

\draw (0,4) node {$\times$};

\draw (0,2.5) node {{\color{red}$\bullet$}};

\draw (0,2.5) node [anchor=east]{{\color{red} $\mathfrak{R}_1 \ni r : p \overset{\beta_1}{\rightarrow} f$}};

\draw (0,1.2) node [anchor =west] {$\left[\begin{tabular}{ccc}
{\color{red}f} & a & n \\ {\color{red}f} & i & k
\end{tabular}\right]$};

\draw (0,1.2) node {$\times$};

\draw (2,-3) node {{\color{blue}$\bullet$}};

\draw (2,-3) node [anchor=east]{{\color{blue}$B^1_e$}};

\draw (2,-1) node [anchor =west] {$\left[\begin{tabular}{ccc}
f & a & n \\ f & i & k
\end{tabular}\right]$};
\draw (2,-1) node {$\times$};

\draw (2,-5) node [anchor =west] {$\left[\begin{tabular}{ccc}
{\color{blue}l} & {\color{blue}i} & {\color{blue}r} \\ f & i & k
\end{tabular}\right]$};
\draw (2,-5) node {$\times$};

\end{tikzpicture}
\end{figure}

 Figure \ref{treeempr} represents these processes and their effect on a small lexicon.

Let $x$ be the position of a creation event. We denote by $U_{x^-}$ and $U_{x^+}$ the value of $U$ just before $x$ and just after $x$, respectively. The creation event changes $U$ as follows:
\begin{equation}\label{eq:creation}
m\sim\mathcal U(1,\ldots,M)\qquad U_{x^+}(m,\cdot)\sim \pi\qquad \forall m'\neq m, U_{x^+}(m',\cdot)=U_{x^-}(m',\cdot).
\end{equation}

Now let $t$ be the position of a transformation event. We draw
\[k\sim\mathcal U(1,\ldots,K)\qquad (a \rightarrow a') = r \sim\mathcal U(\mathfrak{R}_k)\]
so that the transformation is $/a/\to/a'/$ in column $k$. The transformation event changes $U$ as follows:
\begin{eqnarray}\label{eq:transformation}
&\forall k' \neq k, \forall m, &U_{x^+}(m,k') = U_{x^-}(m,k')\nonumber\\
&\forall m, \text{ if } U_{x^-}(m,k)\neq a,  &U_{x^+}(m,k) = U_{x^-}(m,k)\nonumber\\
&\forall m, \text{ if }U_{x^-}(m,k)= a, &U_{x^+}(m,k) = \begin{cases}a' & \text{ w.p. } \beta_k\\a & \text{ w.p. } 1-\beta_k.\end{cases}
\end{eqnarray}

To account for unmodelled events, we allow for noise: at the end of an edge of length $h$, we let
\begin{equation}\label{eq:noise}
\forall m,k, U_{x^+}(m,k) \begin{cases} \sim \mathcal U(1,\ldots,n_k) & \text{ w.p. } 1-e^{-h\nu}\\ = U_{x^-}(m,k) & \text{ otherwise.}\end{cases}
\end{equation}

The noise level $\nu$ is useful to measure model fit. We also rely upon it in the numerical implementation (Section \ref{sec:Numericalmethods}) to define a tempering strategy.

This fully specifies the stochastic process $U$. The data are then observed at the leaves: at  leaf $\ell$, $U_\ell=D_\ell$.

%
%

To summarise, the whole process is described by:

\begin{itemize}
\item $g$ the topology of the phylogenetic tree, with edges $E$.
\item the associated set of edges length $(h_e)_{e \in E}$
\item $\mu$ the renewal rate of the meanings.
\item $\lambda_k$ the rate of the appearance of the transformations for each traits; we write $\bm \lambda=\left(\lambda_k\right)_k$
\item $p_k$ a vector of probabilities for each transformation of each character; we write $\bm p=\left(p_k\right)_k$
\item $\beta_k$ the probability that a transformation is applied; we write $\bm \beta = \left(\beta_k\right)_k$ 
\item $\nu$ the intensity of the noise.
\end{itemize}

\subsection{Questions of interest}

As often in phylogenetic models, the dated tree is the main parameter of interest. However several other question arise once the evolutive model is set. The evolution rates $\lambda$ and $\rho$ in particular, as they inform on evolution speed, can be compared between several subgroups of taxa. 
The parameter $\beta$ is also interesting as it corresponds to different models depending on its value. For $\beta$ close to $1$, the model is close to the classical phonological model that assumes that phonological transformations are \emph{regular}~; by opposition for $\beta$ small, transformations are not regular.

\subsection{Likelihood computation}

\label{sec:cptlkld} 

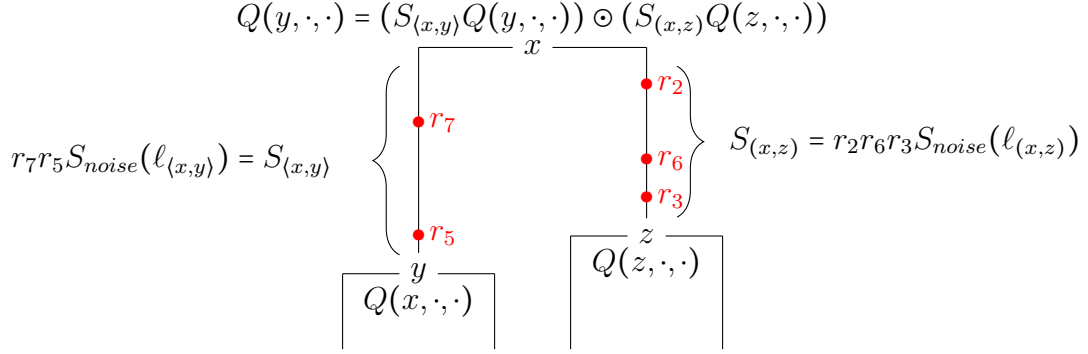
\begin{figure}
\center
\begin{tikzpicture}[scale=.5]

\node (x) at (5,8) {$x$};
\node (y) at (2,2) {$y$};
\node (z) at (8,3) {$z$};

\draw (2,8) -- (x);
\draw (8,8) -- (x);
\draw (0,2) -- (y);
\draw (4,2) -- (y);
\draw (6,3) -- (z);
\draw (10,3) -- (z);

\draw (8,8) -- (z);
\draw (2,8) -- (y);

\draw (10,0) -- (10,3);
\draw (6,0) -- (6,3);
\draw (4,0) -- (4,2);
\draw (0,0) -- (0,2);

\draw (y) node[below] {$Q(x,\cdot ,  \cdot )$};
\draw (x) node[above] {$Q(y,\cdot ,  \cdot ) = (S_{\langle x,y \rangle} Q(y,\cdot,\cdot))\odot (S_{(x,z)} Q(z,\cdot , \cdot))$};
\draw (z) node[below] {$Q(z,\cdot , \cdot )$};

\draw (2,3) node[right,red] {$r_5$};
\draw (2,6) node[right,red] {$r_7$};
\draw (2,3) node[red] {$\bullet$};
\draw (2,6) node[red] {$\bullet$};

\draw (8,7) node[right,red] {$r_2$};
\draw (8,5) node[right,red] {$r_6$};
\draw (8,4) node[right,red] {$r_3$};
\draw (8,7) node[red] {$\bullet$};
\draw (8,5) node[red] {$\bullet$};
\draw (8,4) node[red] {$\bullet$};

\draw [decorate,decoration={brace,amplitude=10pt}] (1.5,2.5) -- (1.5,7.5) node [midway,xshift=-3cm] {$r_7 r_5 S_{noise}(\ell_{\langle x,y \rangle}) = S_{\langle x,y \rangle}$};

\draw [decorate,decoration={brace,mirror,amplitude=10pt}] (8.8,3.5) -- (8.8,7.5) node [midway,xshift=3cm] {$S_{(x,z)} = r_2 r_6 r_3 S_{noise}(\ell_{(x,z)})$};

\end{tikzpicture}
\caption{Computation of the likelihood, without cognacy apparition, where $\odot$ denotes elementwise matrix multiplication.}
\label{fig:lkldcalcul}
\end{figure}

The likelihood is not available in closed form because of the intricacy of the two evolution processes. 
We thus add two sets of latent variables in order to make the computations tractable:

\begin{itemize}
\item $R^k_e$ the list of the dated transformations for character $k$ on branch $e$, with associated position on the branch; we write $\bm R^k=\left(R^k_e\right)_e$
\item $B_e^m$ the position of each apparitions for each meaning $m$ on branch $e$; we write $\bm B=\left(B_e^m\right)_{e,m}$
\end{itemize}

Given these auxiliary variables, the likelihood can be factorized: denoting $\theta=(g,h,\mu,\bm{\lambda},\bm p,\bm\beta,\nu,\bm q) $ the set of all parameters and auxiliary variables, and $D^k=D(\cdot,\cdot,k)$, we have

\begin{eqnarray}
 \mathcal{L}(D \mid \theta,\bm R,\bm B) = \prod_{k} \mathcal{L}(D^k \mid \theta,\bm R^k,\bm B). 
\end{eqnarray}

Then each term $\mathcal{L}(D^k \mid \theta,\bm R^k,\bm B)$ can be computed separately by pruning \citep{Felsenstein}, as follows. In the remainder of this subsection, we fix a character $k$.
The quantity on which we will prune is $\mathbb{P}(D_{desc(x)}(m,k) \mid U_x(m,k)=u,\theta)$ for a node $x$ on the tree, where $desc(x)$ are the descendant leaves of $x$ and $U_x(m,k)$ is the value for character $k$ and meaning $m$.
For the sake of simplicity we drop the dependency in $k$ and $\theta$, and define  $Q(x,m,u)=\mathbb{P}(D_{desc(x)}(m,k) \mid U_x(m,k)=u,\theta)$. 
The likelihood for character $k$ is then:
\[ \mathcal{L}(D^k \mid \theta,\bm R^k,\bm B) = \prod_{i=1}^M \sum_u \pi_0(u) Q(root,m,u). \]

The quantity $Q$ can be computed recursively through branching events of the tree: let $x$ be an internal node with children nodes $y$ and $z$; we have
\begin{eqnarray}
\label{eq:recbranching}
Q(x,m,u) &=&\sum_vQ(y,m,v) P(U_y(m)=v \mid U_x(m)=u) \times \\
&& \qquad  \sum_w Q(z,m,w) P(U_z(m)=w \mid U_x(m)=u).\nonumber
\end{eqnarray}

We are left to compute the quantity $P(U_y(m)=w \mid U_x(m)=u)$ going down a branch $\langle x, y\rangle$. There are two cases to consider for this computation.

\paragraph{Case 1: no cognate apparition} We first consider the case where there is no cognate apparition on branch $\langle x, y\rangle$.
In this case, we have to account for both the transformations and the noise. We define $S$, the transition matrix of the value for a given character, by:

\begin{equation}\label{eq:recnobirth}
S_{\langle x,y\rangle} = \left(\prod_{r \in R_{\langle x,y \rangle}} r  \right)\times S_{noise}(h_{\langle x,y \rangle}), 
\end{equation}
where $S_{noise}(h_{\langle x,y \rangle})$ is the matrix associated with the noise at the end of the edge $\langle x,y \rangle$ from $x$ to $y$, as in eq. \ref{eq:noise}, and where the terms of the product are considered in the order they appear along the branch. In this case, $P(U_y(m)=w \mid U_x(m)=u)$ is equal to the entry at position $(u,w)$ of the matrix $S_\edgexy$.  Figure \ref{fig:lkldcalcul} summarizes the relationship between these objects.

%

\begin{figure}
\center
\begin{tikzpicture}[scale=.5]

\node (x) at (5,8) {$x$};
\node (y) at (2,2) {$y$};
\node (z) at (8,3) {$z$};

\draw (2,8) -- (x);
\draw (8,8) -- (x);
\draw (0,2) -- (y);
\draw (4,2) -- (y);
\draw (6,3) -- (z);
\draw (10,3) -- (z);

\draw (8,8) -- (z);
\draw (2,8) -- (y);

\draw (10,0) -- (10,3);
\draw (6,0) -- (6,3);
\draw (4,0) -- (4,2);
\draw (0,0) -- (0,2);

\draw (y) node[below] {$Q(\cdot , y, \cdot )$};
\draw (x) node[above] {$Q(x,m^,u) = \sum_v \sum_w \left[\pi_{\langle x,y \rangle,\tau}(v) Q(m,y,v) \right] Q(m,z,w)S_{(x,z)}(u,w)$};
\draw (z) node[below] {$Q(\cdot ,z, \cdot )$};

\draw (2,3) node[right,red] {$r_5$};
\draw (2,6) node[right,red] {$r_7$};
\draw (2,3) node[red] {$\bullet$};
\draw (2,6) node[red] {$\bullet$};

\draw (2,4.5) node[left] {$\pi_{\langle x,y \rangle,\tau} = \pi_0 r_5  S_{noise}(\ell_{\langle x,y \rangle})$};
\draw (2,4.5) node {$\bullet$};

\draw (8,5.5) node[right] {$S_{(x,z)}$};

\draw[<->] (1.5,2.5)--(1.5,4);
\draw (1.5,3.25) node[left] {$\tau$};


\end{tikzpicture}
\caption{Computation of the likelihood for meaning $m$, with an apparition on the edge $\langle x,y \rangle$.}
\label{fig:lkldcalcul2}
\end{figure}

\paragraph{Case 2: at least one cognate apparition}
The second case to consider is if there is apparition of a cognate on the edge $\langle x,y \rangle$ for meaning $m$.
In this case, the value at the child node $U_y(m)$ is independent of the value at the parent node $U_x(m)$.
Let $\tau$ be the last apparition time of a cognate for meaning $m$ on branch $\edgexy$. The distribution for character $k$ of meaning $m$ at the end of the branch $y$ only depends on the transformations $\{r \in R_{\edgexy}^k:t(r) < \tau\}$, and is given by:

\begin{equation}\label{eq:recwithbirth}
 \pi_{\edgexy,\tau}=
\pi_0 \times \left( \prod_{r \in R_{\edgexy}^k, \ t(r)<\tau} r \right) \times S_{noise}(h_{\edgexy}) 
\end{equation}

and $P(U_y(m)=v \mid U_x(m)=u) $ is equal to the $v$th entry of vector $\pi_{\edgexy,\tau}$.  We represent these computations in Figure \ref{fig:lkldcalcul2}.


\bigskip

We initialize $Q$ at the leaves:  $Q(x,m,u) = \mathbf{1}_{D_x(m)=u} $ if $x$ is a leaf.
Taken together, Equations \ref{eq:recnobirth} and \ref{eq:recwithbirth} allow us to compute the quantity $Q$ on which we prune in equation \ref{eq:recbranching}. We are thus able to calculate the likelihood.

\subsection{Missing data}

The computations above assume that all data have been observed. 
Our model offers the advantage of handling missing directly, without any more cost.
The matrix $Q$ at a leaf $\ell$ is initialized with the observations; it represents the probability, knowing the state at the leaf to observe the observations at this leaf. In the case of missing data, since the pattern of missingness is observed, it is straightforward to modify the leaf initializations as

\begin{align*}
Q(\ell,m,k) &= P(D_\ell(,m,k) \mid U_\ell(m,k),\text{missingness}) \\
&= \left\lbrace \begin{matrix}
1 & \text{ if missing} \\
0 & \text{ if not missing and } D_\ell(m,k)\neq U_\ell(m,k) \\
1 & \text{ if not missing and } D_\ell(m,k)=U_\ell(m,k) \end{matrix} \right.
\end{align*} 


\subsection{Prior distributions}

\subsubsection{On the evolution parameters}

We take conjugate prior distributions on the evolution parameters. For  $\lambda$, $\mu$ and $\nu$ we chose gamma priors fitted so that on each branch there are, in expectation, less than ten transformations or apparition, in order to keep the model in reasonable.

We use conjugate priors for $\lambda$, $\mu$, $\nu$ and $p$ as these are the parameters associated with Poisson and multinomial distributions. In our applications below, we take priors $\beta_k\sim Beta(1, 10)$ iid and $\pi(\nu)\propto 1/\nu \mathbf{1}_{[a,b]}(\nu)$.

\subsubsection{On the tree}

Following \cite{Nicholls2007dated}, we use a prior on the trees of the form:

\[ \pi(g) \propto \mathbf{1}_{t(root) \in [a,b]} t(root)^{L}, \]

where $L$ is the number of leaves and $a$ and $b$ are bounds on the root age. We further constraints this prior by adding constraints on clades and their ages. Notice that learning the root age is only possible if enough internal constraints are provided, knowing this, we also propose to fix the root age to an arbitrary value, so that we can focus on learning topologies and relative ages --- which does not depend on the root age. As appealing as learning the root age is, we advise against putting too much faith in the posterior~; in general this age is little informed, resulting only in the propagation of a few constraints on the leaves. This was, anyway not a primary concern of our study.

We also tried to add Gamma prior on the total length of the tree, that is
\[ \pi(g) \propto \Gamma_{c,d}\left( length(g) \right) \mathbf{1}_{t(root) \in [a,b]} t(root)^{L}, \]

where $\Gamma_{c,d}$ is the density of a Gamma distribution with parameters $c$ and $d$. The results where similar compared to the uniform prior, which indicates a low effect of the prior.

\subsection{Inference of ancestral states}

As stated in Section \ref{sec:cptlkld} the matrix $Q$ contains the probability of the observations given the state of each character for each meaning. Given the latent variables $R$ and $L$, it is possible to compute at any node $x$ of the tree $\pi_x^k(m)$ the distribution of the values for meaning $m$, and character $k$ at $x$, by descending the tree from the root, at which $\pi_r^k(m)=\pi_0$, to $x$. We can thus compute exactly, with $n_k$ the number of possible values for character $k$: \[\pi(U(x,m,k) \mid R,L,D) = \frac{Q(x,m,k) \odot \pi_x^k(m)}{\sum_{i=1}^{n_k} Q(x,m,k)(i)\pi_x^k(m)(i)},\] the renormalization is computed easily as a sum over a dozen elements. Notice that in this case, when conditioning on $R$ and $L$, the value for each meaning and each character is independent. If we are only interested in the distribution of the ancestral value for a given meaning and character, this is of little interest, but if we are interested in the \emph{joint} distribution of several characters and meanings, care must be taken when renormalizing the conditional posteriors.

The posterior distribution is then:
\[ \int_{R,L} \pi(U_x(m,k) \mid R,L,D) \pi(R,L \mid D) \mathrm{d}(R,L), \]
which can be approximated by Monte Carlo, as we already have a sample from $\pi(R,L \mid D)$.


\section{Numerical methods}
\label{sec:Numericalmethods}

All the code is available, along with the datasets used, at \url{https://github.com/GClarte/PhylogenyFromMatrices}. 

We use a Sequential Monte Carlo (SMC, \cite{del2006sequential}) algorithm to sample from the posterior. SMC relies on importance sampling principles.  This method requires us to define a sequence of distributions $\pi_t$ that  creates a smooth transition between an initial known distribution and the posterior. The usual method is to use a tempering of the form $\pi_t(x) \propto \pi(x \mid D)^{1/t} $ or $\pi_t(x) \propto \pi(x) L(D\mid x)^t $. 
This strategy is impractical in our case, as our mutation kernel makes use of a Gibbs sampler, disallowing such tempered intermediate distributions; none of our attempts in this direction were fruitful.
Instead, we use the noise parameter as a tempering parameter. For a high value of $\nu$, on each branch the noise matrix $S_{noise}$ is close to the matrix $ (\pi_0, \dots, \pi_0) $, that is as if the model was a forest of trees constituted of only one leaf. Conversely, for $\nu$ small we retrieve the model we are looking for.

As the noise is a parameter that still needs to be learned, we propose the following distributions. For  a sequence $\nu_1 > \dots > \nu_T$ we define :

\[ \pi_t(g,\mu,\lambda,p,\beta,q, \nu \mid D) = \pi( g,\mu,\lambda,p,\beta,q \mid \nu = \nu_t, D), \]

with a final target

\[ \pi_{T+1}(g,\mu,\lambda,p,\beta,q, \nu \mid D) = \pi( g,\mu,\lambda,p,\beta,q, \nu \mid D). \]

That is we fix the noise parameter and we reduce it until it reaches a value that will lie in a part of the space with non negligible posterior probability. 

This does not ensure a proper mixing of the SMC algorithm, as the number of tempering steps needed can be quite high --- and a high number of steps would require a higher number of particles, which becomes quite heavy in terms of memory costs --- leading to a genealogy degeneracy. This is a severe limitation to the use of our current version of the method, and this prevents us from estimating the marginal likelihood of the model. Depsite the particle degeneracy, we notice that the last particles do mix adequately. Our chosen solution is thus to start with an SMC run, and use the final step to initialize a Hastings-Metropolis-Rosenbluth-Teller algorithm \citep{Metropolis1953Equation, hastings1970monte}. We describe thoroughly in the supplementary the proposal kernel we used, the main difficulty lying in the handling of the latent variables.


This method allows for 
parallelization, but it is at the cost of memory efficiency, because of the size of the latent variables: 
we have to store the value of $Q$ at each node (that is, a matrix of size $n_k \times M$ with $n_k$ number of possible values for character $k$ and $M$ number of cognates), for each character. In particular the number of particles is problematic. On simulated data, 2000 particles are enough to provide stable results. On real dataset, with larger inventories for each character the number of particles cannot be higher than 1500, and the results are sometimes unstable. We are aware of this issue, and have for now no other solution than increasing the available memory.

\subsection{Rooting the tree}

This SMC scheme is able to adequately explore the posterior; with one exception. Analysis of the output of the initial implementation of this algorithm showed that the split closest to the root (\emph{i.e.} the choice of outgroup) mixed poorly in several examples. To circumvent this issue, 
we added another taxon to the dataset with only missing data, constraining this taxon to be an outgroup. The results following this method have been satisfying, although we note that this a consequence of this choice is a small change in the prior on the MRCA associated with the initial languages, that is the root of the studied taxa.

\section{Applications}
\label{sec:appli}

To demonstrate the applicability of our method, we first apply it to synthetic datasets in Section \ref{sec:ling:synthetictest}, and show that we are able to correctly reconstruct parameters of interest.  We then apply it to a subset of Polynesian languages in \ref{sec:polynesian}.

We point out that our model is designed for particular datasets which present the desired alignment properties. 
For many language families, this alignment is not verified, and so our model would only be applicable with difficulty. Nonetheless, there are numerous cases that are of interest. For example, many Tai languages, which present monosyllabic structure, correspond to our framework. Below, we apply our method to certain meanings from a set of well-chosen Polysenian languages, which all fit a  CVCV pattern.

Noticeably, these examples corresponds to closely related languages, that often have only a few cognacy classes to differ, that is a very faint cognacy signal. In such cases, using the phonological signal is essential. Additionally, these data allow for an additional check: it is possible to check the cognacy reconstruction by comparing with the linguists' reconstruction.



\subsection{On synthetic data}

\label{sec:ling:synthetictest}

First, we can run numerical methods on synthetic datasets. Some experiments with data from misspecified models are available in supplementary material.

For this experiment, we chose 4 characters with the following specifications : 

\hspace{-1cm}\begin{tabular}{c|c|c|c|c|c}
Number of value & Possible transitions & $\lambda$ & $\pi_0$ & $p$ & $\beta$ \\
\hline
 5 & $1 \leftrightarrow 2 \leftrightarrow 3 \leftrightarrow 4 \leftrightarrow 5 \leftrightarrow 1 \leftrightarrow 4$ & $10^{-2}$ & $\mathcal{U}\lbrace 1, \dots, 5 \rbrace$ & uniform & 1\\
 \hline
 5 & $1 \leftrightarrow 2 \leftrightarrow 3 \leftrightarrow 4 \leftrightarrow 5 \leftrightarrow 1 \leftrightarrow 4$ & $10^{-2}$ & $\mathcal{U}\lbrace 1, \dots, 5 \rbrace$ & uniform & 1\\
 \hline
  2 & $1 \leftrightarrow 2$ & $10^{-3}$ & $\mathcal{U}\lbrace 1, 2 \rbrace$ & uniform & 1\\
  \hline
   2 & $1 \leftrightarrow 2$ & $10^{-3}$ & $\mathcal{U}\lbrace 1, 2 \rbrace$ & uniform & 1
\end{tabular}

\vspace{0.5cm}

\begin{figure}
\centering
\includegraphics[scale=.4]{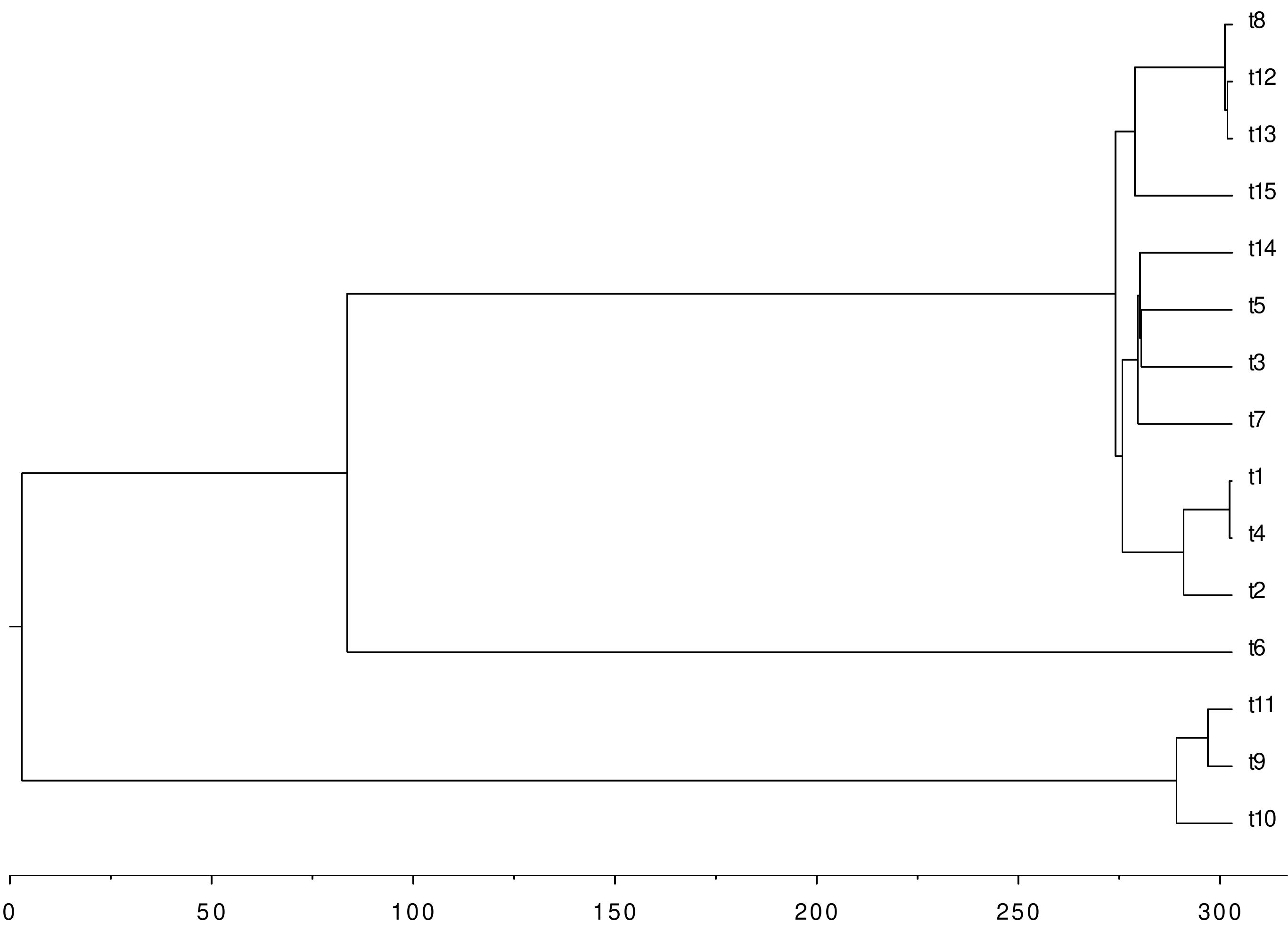} 
\includegraphics[scale=.4]{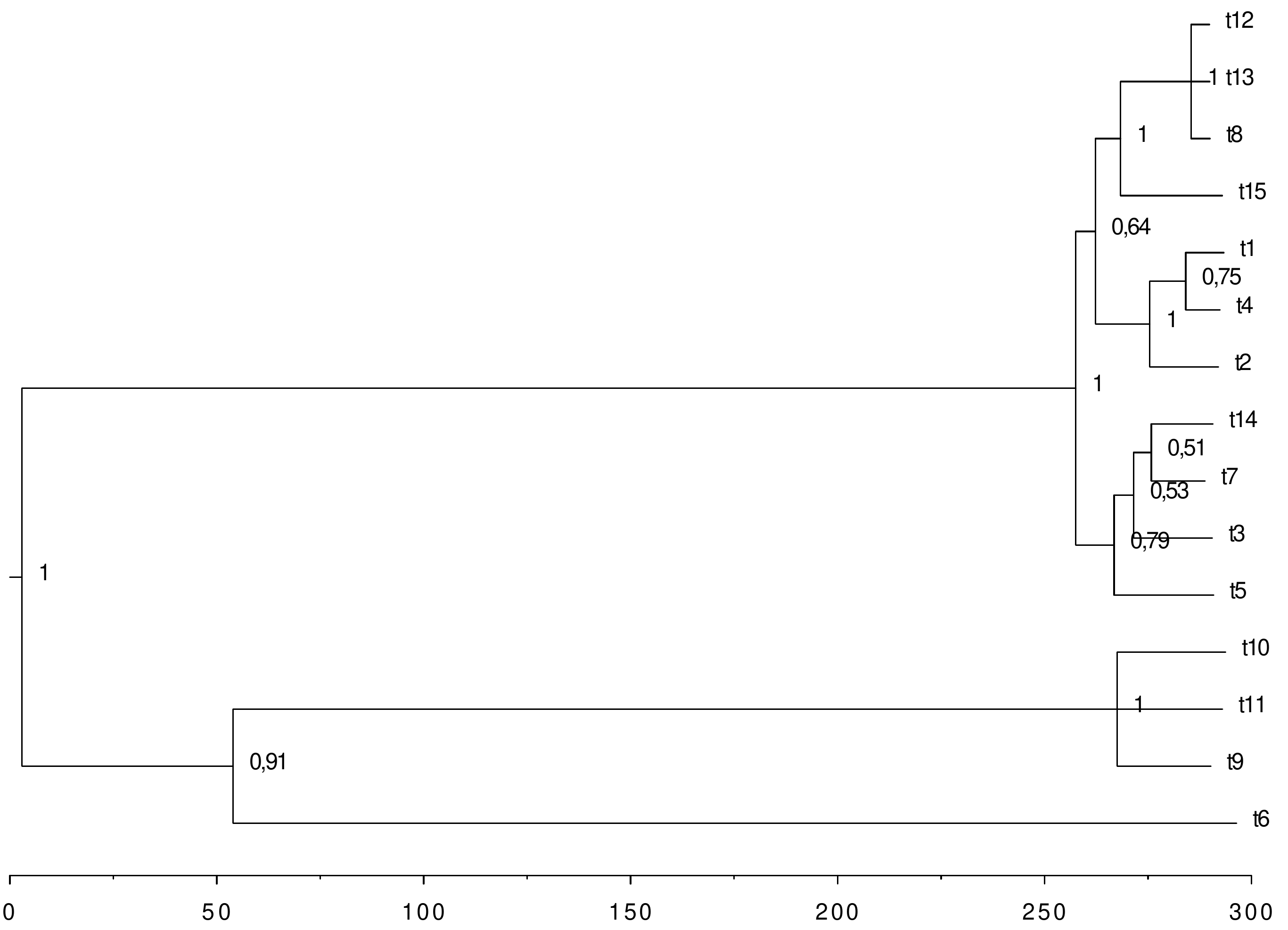} 
\caption{True tree and consensus tree of the final sample, simulated dataset. Internal nodes of the true tree are tagged with their posterior probability.}
\label{fig:arbresimu15}
\end{figure}

The true tree is represented in top of Figure \ref{fig:arbresimu15}. We constrain leaves $\{8,12,13,15 \}$ and $\{9,11,10\}$ to form clades and we added the following constraints on some internal node ages: the nearest common ancestor of $10$ and $9$ in $[10,200]$, of $1$ and $2$ in $[10,200]$, of $8$ and $15$ in $[20,30]$. Different constraints on the internal nodes led to very similar results.

We generated $100$ data points with these characters on a $15$-leaved tree presented on the left in Figure \ref{fig:arbresimu15}. We ran the previously described SMC sampler with a twelve step tempering on $\nu$ : $ \{ 1\cdot10^{-1}, \ 5\cdot10^{-2}, \ 3\cdot10^{-2}, \ 2\cdot10^{-2}, \ 1\cdot10^{-2}, \ 9\cdot10^{-3}, \ 8\cdot10^{-3}, \ 7\cdot10^{-3}, \ 6\cdot10^{-3}, \ 5\cdot10^{-3},\ 1\cdot10^{-3} \}$. We used $2\cdot 10^3$ particles, for the mutation step we ran $10^3$ iterations of our Metropolis within Gibbs algorithm. After the last tempering, we ran a last mutation step of $10^4$ iterations for each particles. In total, on $30$ CPU Intel Xeon CPU E5-2630 v4 \@ 2.20GHz, it took roughly 5 hours.

The consensus tree is represented on the right in Figure \ref{fig:arbresimu15}. The overall topology is well reconstructed, with uncertainty on the smaller branches, which is expected as the small branches rarely host even a single transformation. Noticeably each group of leaves supported by the consensus tree corresponds to a true clade. Node ages do not seem accurately reconstructed, in accordance with the root being to high.

The parameters are also well reconstructed; details are given in Supplementary material. The existence of apparition events on the path between two leaves is also quite well reconstructed, which has an interesting interpretation.

%

\subsection{Polynesian languages}
\label{sec:polynesian}

We now present a preliminary application of this method to data for 78 meanings from 14 Polynesian languages. The raw data form a subset of the online Austronesian Basic Vocabulary Database (ABVD) \citep{abvd}. We selected a subset of languages and meanings for which many word forms were 4 characters long, of the form Consonant-Vowel-Consonant-Vowel, to allow for perfect alignment. This preprocessing step led us to exclude from the data all words which are not of length 4 and replace them by missing data. This step biases the data, and we therefore do not claim the inferred topology is reliable. The cognacy judgments will also be biased, but we might hope nonetheless that the inferred cognacies are somewhat trustworthy. 

\begin{figure}
\includegraphics[scale=1]{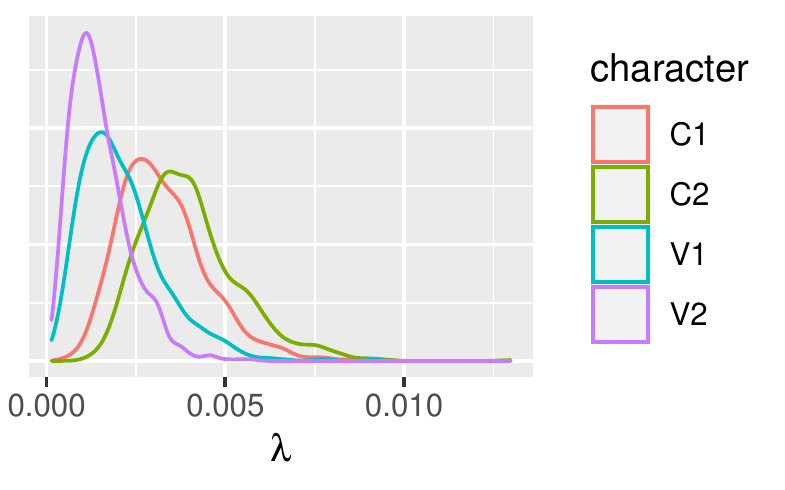}
\includegraphics[scale=1]{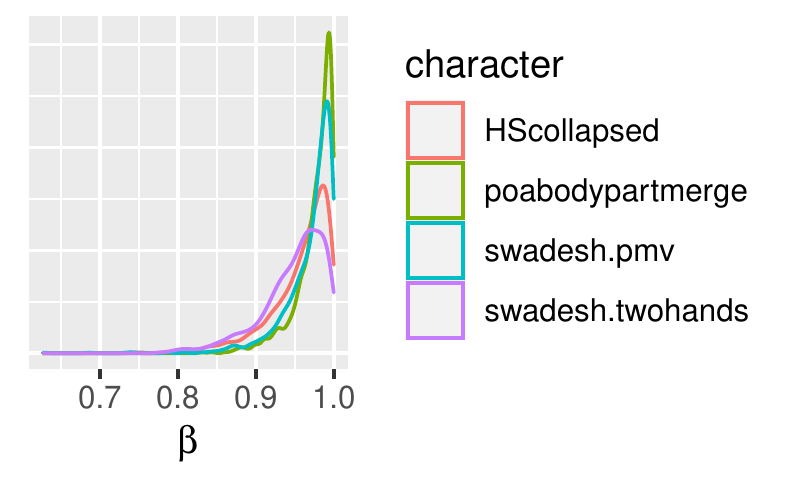}
\includegraphics[scale=1]{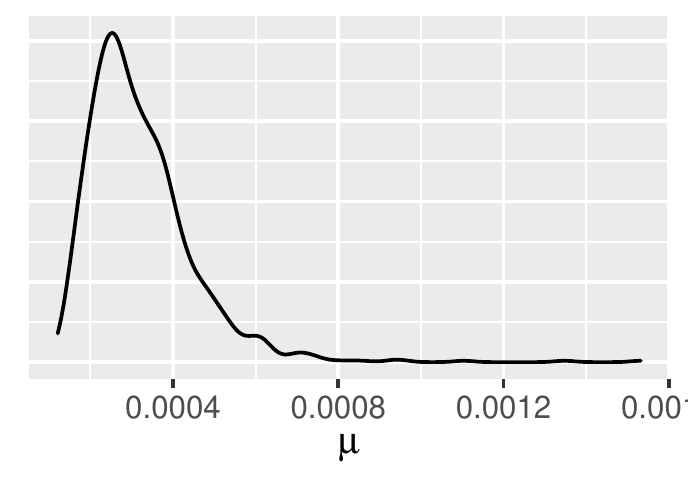}
\includegraphics[scale=1]{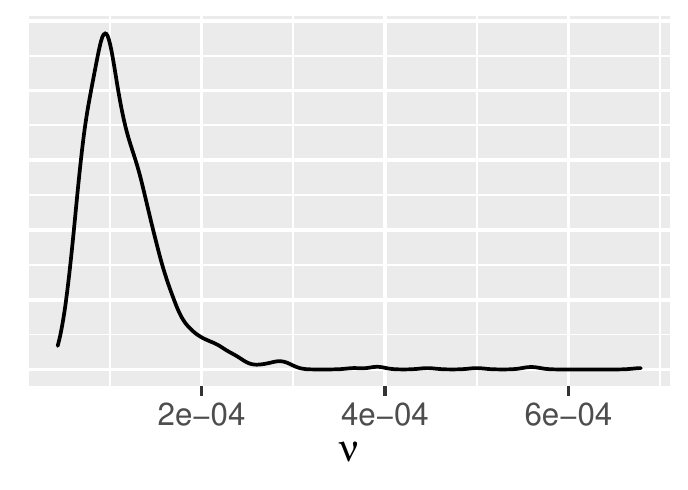}
\caption{Parameter estimations from the ABVD dataset}
\label{fig:parampoly}
\end{figure}

\begin{figure}
\includegraphics[scale=0.5]{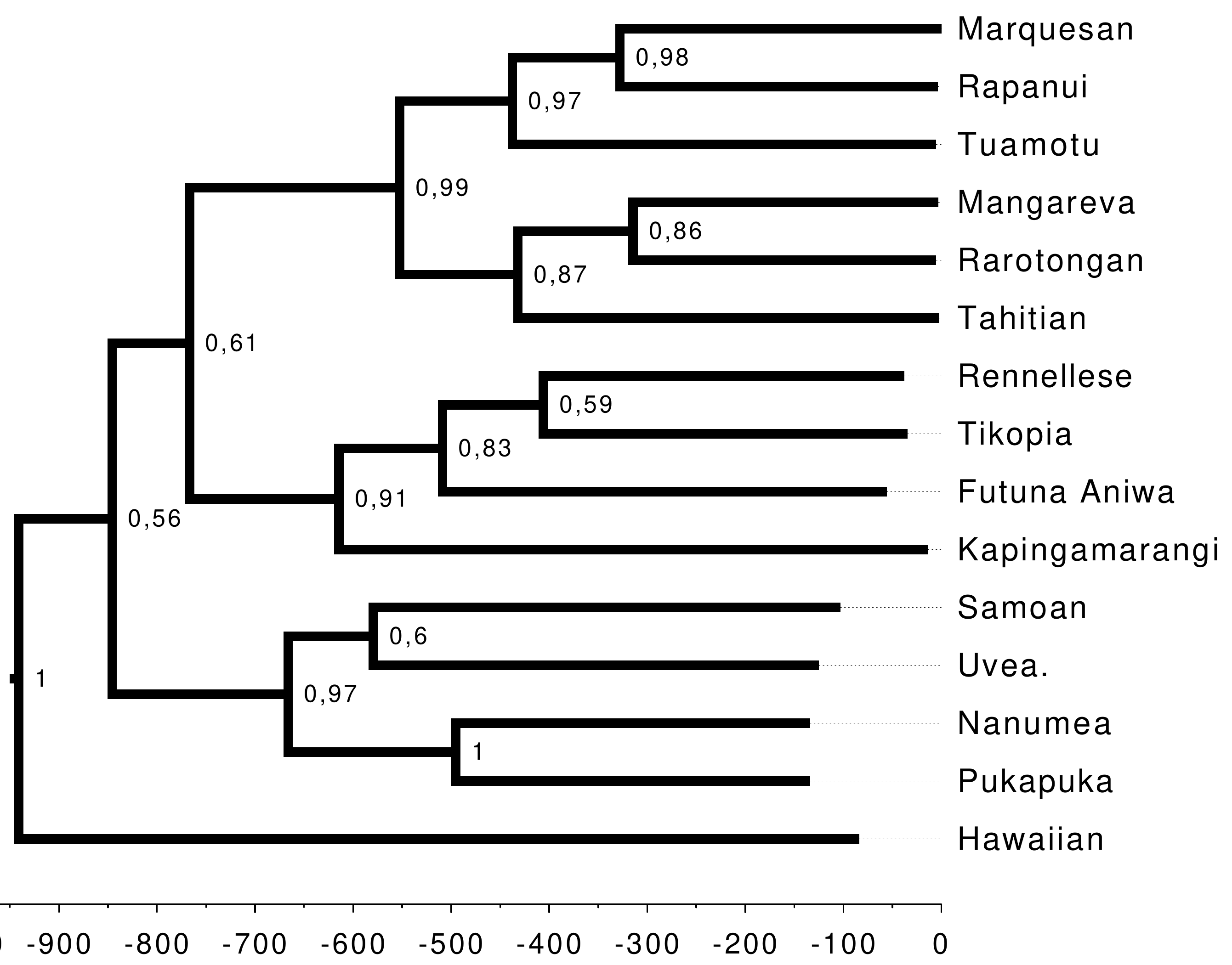}
\caption{Consensus tree from the ABVD dataset}
\label{fig:topopoly}
\end{figure}

Despite this caveat, the results are interesting. In particular, the evolution parameters in Figure \ref{fig:parampoly} show a similarity between the evolution rate of the two vowels, and between the two consonants. The noise parameter is relatively low, indicating that the model is indeed detecting a signal from the dataset. The topology is summarized by a majority-rule consensus tree in Figure \ref{fig:topopoly}). The position of Hawaiian as an outlier in 54\% of the posterior is unexpected, and might be due to a bias in the way we selected the data. The other Eastern Polynesian languages (Marquesan, Mangareva, Tuamotu, Rarotongan, Tahitian and Rapanui) form a clade with posterior probability 99\%, separating from the other languages (which correspond to the Samoic languages). The Futunic languages (Rennellese, Tikopia and Futuna Aniwa) also form a strongly supported clade.

\begin{figure}
\centering
\includegraphics[width=15cm]{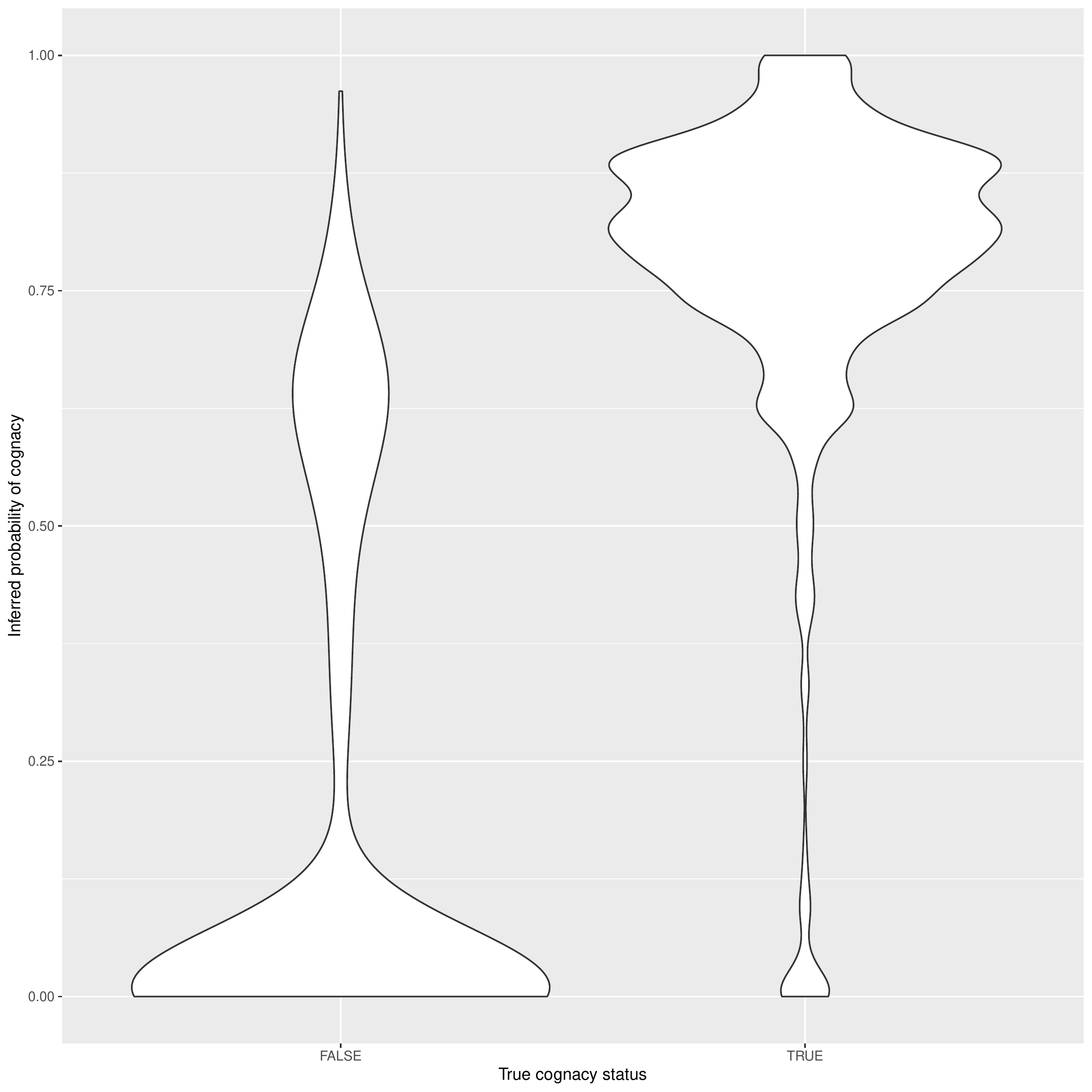}
\caption{Estimated probability of cognacy for pairs of words in the ABVD dataset. Left: pairs of words which are not cognates according to expert linguists; right: pairs of words which are true cognates. \label{fig:checkcog}}
\end{figure}

The ABVD includes cognate sets by expert linguists for most pairs of words, so we compare the inferred probability of cognacy with the true cognacy status. There is overall agreement between the inferred and true cognacy status of pairs of words, as  shown in Figure \ref{fig:checkcog}. If we use the decision rule that two words are inferred to be cognates iff the probability of cognacy is $>0.5$, we correctly classify the cognacy status of 88\% of word pairs.

In ongoing work not presented in this preprint, we apply this method to other language families which correspond better to our alignment requirement.

\section{Discussion}

This works presents a new step towards an automatic version of the quantitative method, which is arguably one of the ultimate goals of computational phylolinguistics: learning cognacy classes along with the phylogeny and evolution parameters. 

 This model is of course an imperfect representation of the true evolution of lexicon and phonology. The first glaring simplification is that we do not allow for the phonological transformations to depend on context. (Instead, each transformation applies with probability $\beta_k\in ]0,1[$, which allows for transformations to not apply to all words.)
 


On the numerical side, the memory cost of our method --- that have been increased to decimate the time cost --- remains the main limitation, computing the marginal likelihood of our model is also out of reach for now.

An attainable extension of our model   would be to allow for some or all cognacy judgments to be known, constraining the latent variables $\bm B$, but still inferring the phonological transformation to better inform the phylogeny.
Further down the road, future work will need to address modelling context-dependent phonological changes, as well as insertions and deletions of phonemes, and then inferring under such a model in an efficient manner.  Nonetheless, we believe this work to be an important step towards this ultimate goal, and to already allow inference of language histories for lesser-studied families for which no cognate data is available.

\bibliographystyle{plainnat}
\bibliography{mabiblio}

\newpage

\part{Supplementary Materials}

\subsection{Metropolis-Hastings steps}

For the mutation steps, we run many iterations of a simple Metropolis-Hastings within Gibbs sampler. That is we update one by one each coordinate of the parameter and latent variables conditionally on the value of the data and the other coordinates. More precisely we adopt the following update schemes, the probability to choose each of these update is chosen by the user so that the most important and most difficult updates are the most frequent. More precisely, we favour the update of the latent variables $R$ and $B$, and the topology $g$.

We write $L$ the full likelihood computed as $\sum_m\pi_0U_{root}(m,\cdot)$, where $U_{root}$ is computed through pruning as described in the main text.

\subsubsection{Update of $R$}

We sample the value of the transformations from the prior. We also tried to add or remove one by one the transformations but that lead to poor acceptance rate, this is due to the high correlation of the transformations, especially if all the transformations between two values do not exist. The prior on the transformations is induced by $k$, $p$ and $\lambda$.

The conditional writes \[\pi(R^k \mid B,\lambda, \mu, p, \beta, \nu,D) \propto L(D \mid \theta)\frac{(\lambda h_e )^{\vert R \vert}}{\vert R \vert!} p^R,\] 
where $p^R = \prod_{r \in R} p[r] $ and $\vert R \vert= \sum_e \vert R_e \vert$.

The density of the kernel cancels out the term $\frac{(\lambda h_e )^{\vert R \vert}}{\vert R \vert!} p^R$ in the Metropolis acceptance ratio.

\subsubsection{Update of $B$}

We choose an edge and we change the whole vector of apparition for each cognate, we propose a new value of $B$ from the prior and accept it independently. This may seem too simple, but it provides better results than the addition and removal of single transformation in $B$ --- as the transformations are highly correlated.

As previously, the proposal density cancels out a part of the acceptance ratio, only leaving the likelihood ratio.

Noticeably for this latent variable, we did not store the position of each apparition but only the total number and the position of the last, which makes the update arguably more difficult in the SPR moves.

\subsubsection{Update of the parameters $\lambda$, $\mu$, $p$, $\beta$ and $\nu$}

We chose, when possible, conjugated priors --- Poisson and multinomial distributions. This makes the update of the parameters quite simple. The other parameters are updated according to a Metropolis-Hastings step.

For $\lambda_k$, the update is exact, from the conjugated posterior $\Gamma(\alpha + n, \beta + l)$, where $n$ is the number of transformations on the whole tree for character $k$, and $l$ the length of the tree.

For $p$ the update is also exact, as the conditional is merely $\mathcal{D}ir(\alpha + \hat{R}) $, where $\hat{R}$ is the vector of counts of the nature of the transformations.

For the other parameters, we rely on a Metropolis-Hastings step, with a proposal from the prior for $\beta$, and Gaussian for $\nu$ and $\mu$. An exact sampling could be use for $\mu$ but we tried non conjugate priors as well.

\subsubsection{Update of $t$}

Concerning the update of the length of the edges, we use a simple MH step, where we propose to move a node in the possible range of time constrained by the parent and children nodes. We propose also to rescale the tree, or part of the tree. If the root age is fixed, some of those moves are impossible.

The conditional writes:

\[ \pi(t \mid D,B,R,\lambda,\rho,\nu) \propto L(D \mid D,B,R,t,\nu)\prod_{k,e}\frac{(\lambda_k h_e )^{\vert R^k \vert}}{\vert R^k \vert!}\frac{(\mu h_e )^{\vert B \vert}}{\vert B \vert!}, \]

the proposal is symmetric which cancels the jacobian.

\subsubsection{Update of the topology}

This last move is the most important and consequently the most difficult to choose. In the previous works \citep{GrayRussellAtkinsonQuentin,Ryderdphil} a new tree is built by an SPR (sample-prune-regraft) move, in practice a subtree is always regraft at the nearest possible position --- its cousin, meaning that the (sub)-tree $((a,b),c)$ becomes $(a,(b,c))$. 

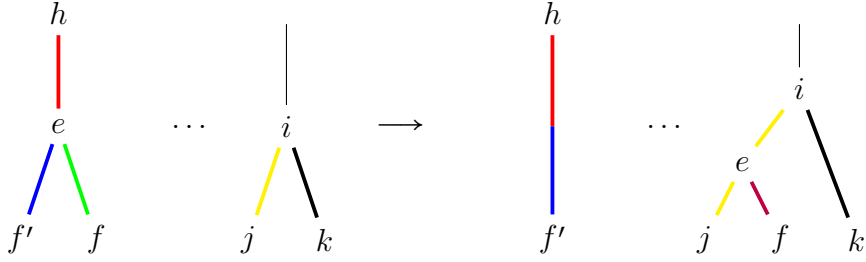
\begin{figure}
\center
\begin{tikzpicture}[scale=.5]

\node (h) at (1,6) {$h$};
\node (e) at (1,3) {$e$};
\node (f) at (0, 0)  {$f'$};
\node (g) at (2,0) {$f$};
\node (ii) at (7,6) {};
\node (i) at (7,3) {$i$};
\node (j) at (6,0) {$j$};
\node (k) at (8,0) {$k$};

\node (h') at (14,6) {$h$};
\node (e') at (19,2) {$e$};
\node (f') at (14,0)  {$f'$};
\node (g') at (20,0) {$f$};
\node (ii') at (20.5,6) {};
\node (i') at (20.5,4) {$i$};
\node (j') at ( 18,0) {$j$};
\node (k') at (22,0) {$k$};

\draw [line width=0.05cm,red](h) -- (e);
\draw [line width=0.05cm,blue](e) -- (f);
\draw [line width=0.05cm,green](e) -- (g);
\draw (ii) -- (i);
\draw [line width=0.05cm,yellow](i) -- (j);
\draw [line width=0.05cm](i) -- (k);

\draw [line width=0.05cm,red](h') -- (14,3);
\draw [line width=0.05cm,blue](14,3) -- (f');
\draw (ii') -- (i');
\draw [line width=0.05cm,yellow](i') -- (e');
\draw [line width=0.05cm,yellow](e') -- (j');
\draw [line width=0.05cm,purple](e') -- (g');
\draw [line width=0.05cm](i') -- (k');

\node at (4.5,3) {$\dots$};
\node at (10,3) {$\longrightarrow$};
\node at (17,3) {$\dots$};

\end{tikzpicture}
\caption{Move of the subtree starting at $f$ so that $f$ has $j$ as brother. The colors represent the edge-attached latent variables.}
\label{fig:exmovetree}
\end{figure}

The question that remains is how to transport the latent variables of the changed parts of the tree. In order to preserve the intuition behind the latent variables, it seems reasonable to either keep all the ages of the nodes constant, or to change the ages while only moving to the nearest possible position --- it is not possible to keep both as it would be incompatible most of the time. We describe in algorithm \ref{alg:SPR1} the main method we used, a graphical representation can be found in Figure \ref{fig:exmovetree}, the case of change with the nearest neighbour corresponds to $i=h$.

\begin{algorithm}[H]
\SetAlgoLined
Let $g$ be a tree, $X$ the edges dependent latent variables.

Choose $f \in g$ a non root node, let $e$ be its parent, $h$ its grandparent and $f'$ its brother.

Choose $j$ a node, whose parent is denoted $i$, such that $t(e) \in [t(j),t(i)]$ where $t$ is the height of  node.

Set $j$ as the new brother of $f$, by splitting the edge $\langle i,j \rangle$ into $\langle i, e \rangle$ and $\langle e,f\rangle$, such that $t(e)$ remains constant. Merge the edges $\langle h,e \rangle$ and $\langle e,f'\rangle$.

Merge the latent variables on $\langle h,e \rangle$ and $\langle e,f' \rangle$ into latent variables on $\langle h,f' \rangle$.

Split the latent variables on $\langle i,j \rangle$ into latent variables on $\langle i,e \rangle$ and $\langle e,j \rangle$ proportionally to the length of the edges.

Sample from the prior latent variables on $\langle e,f \rangle$.

\caption{Sample Prune and Regraft algorithm with latent variables. See Figure \ref{fig:exmovetree}.}
\label{alg:SPR1}

\end{algorithm}

\subsection{Further results on the simulated dataset}

\begin{figure}
\center
\includegraphics[scale=.9]{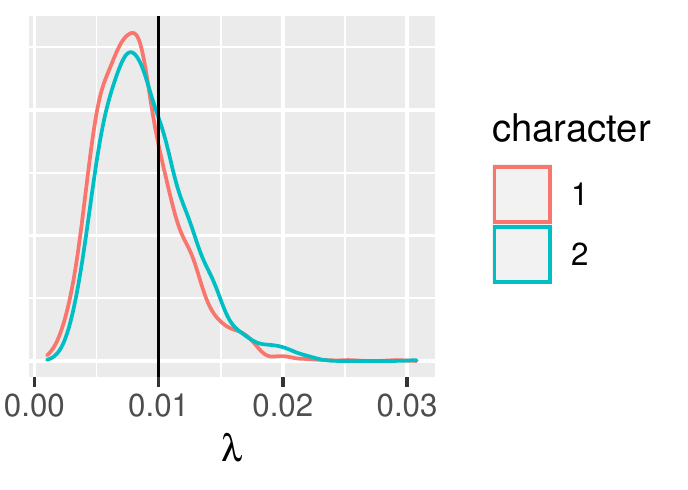} \includegraphics[scale=.9]{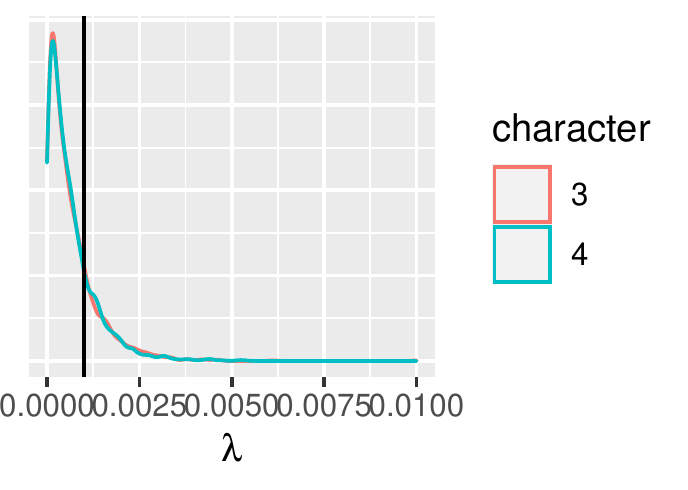} \includegraphics[scale=.9]{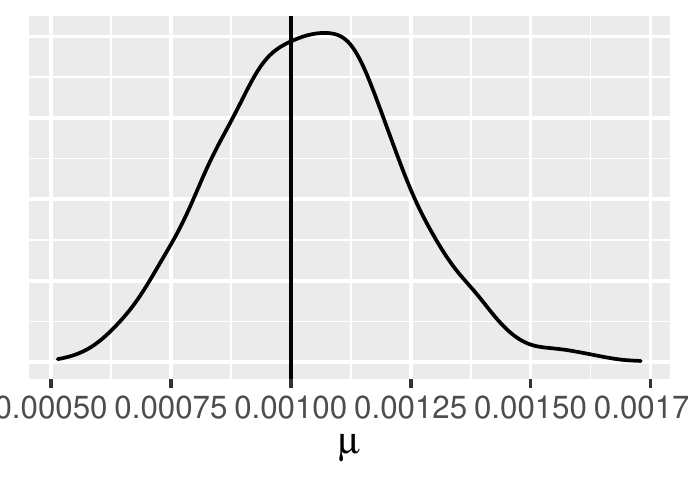}  \includegraphics[scale=.9]{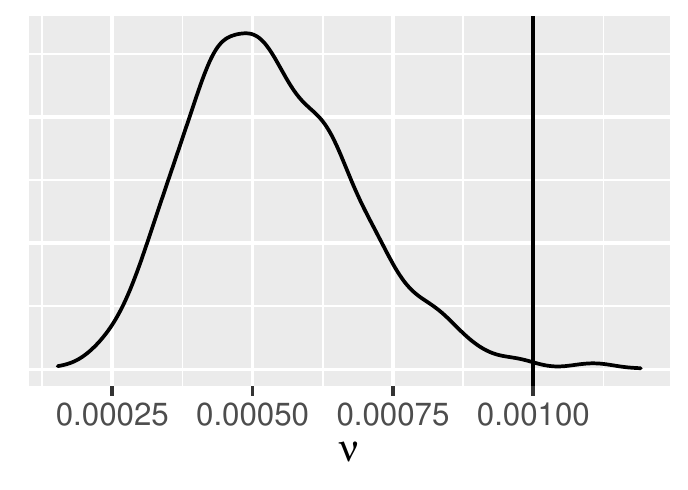} \includegraphics[scale=.9]{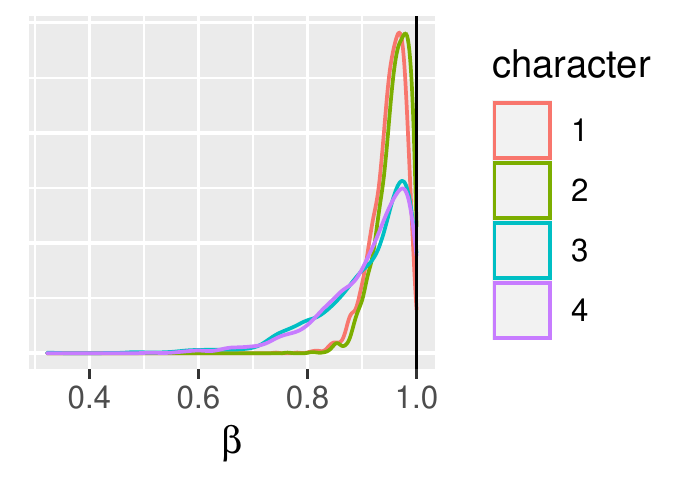} \includegraphics[scale=.9]{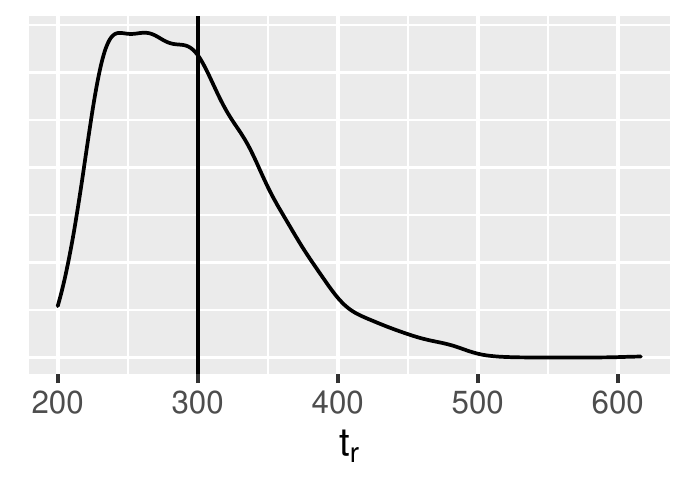}
\caption{Posterior estimations of $\lambda$, $\rho$, $\nu$, $\beta$ ant the root age. The true values are indicated by the vertical lines.}
\label{fig:simusparam15}
\end{figure}

\begin{figure}
\center
\includegraphics[scale=.9]{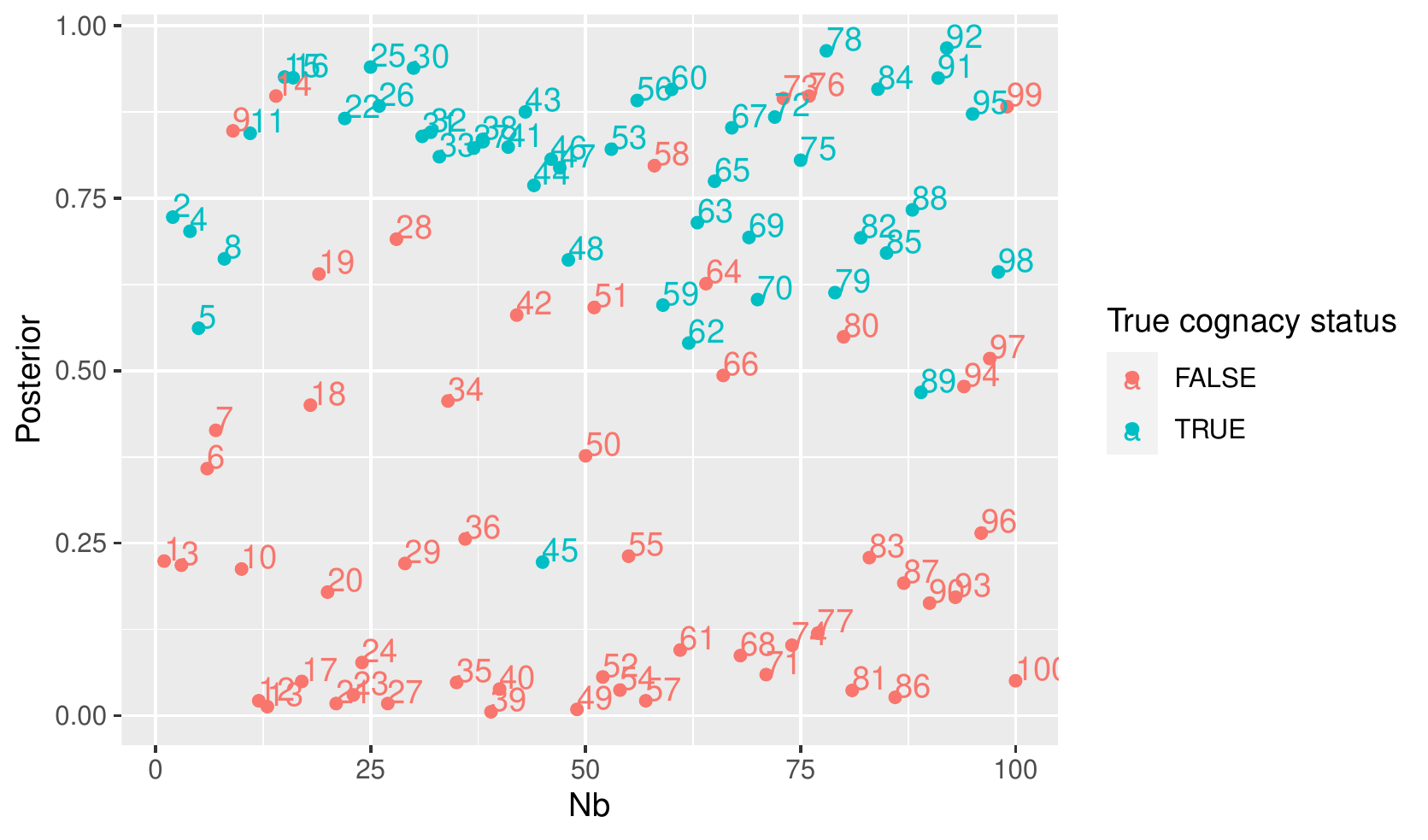} 
\caption{Posterior cognacy probability for each meaning between $t4$ and $t6$. The color indicates the true cognacy status.}
\label{fig:simuscog15}
\end{figure}

On the side of the parameters, we provide in Figure \ref{fig:simusparam15} the posterior distributions for the main evolution parameters. They seem quite efficiently reconstructed ---~particularly the transformation rate $\beta$ which is close to $1$, the true value~--- with an underestimation overall the edge length parameters, which is linked to the overestimation in edge length ---~the number of events per branch is well reconstructed in the lower parts of the tree~---. This induces a poor performance in the reconstruction of cognacy when studying distantly related leaves, as the number of apparitions on the long edges starting from the root is underestimated. Maybe the number of character should be increased so that the creation of new words becomes \og cheaper \fg{} compared to the evolution. In Figure \ref{fig:simuscog15} we represent the posterior cognacy probability. The number of false positive is quite high. However, in general, meanings with low cognacy probability do corresponds to words that are non cognates in truth. The root age is overestimated in accordance to what has been said before.

\begin{figure}
\center
\includegraphics[scale=1]{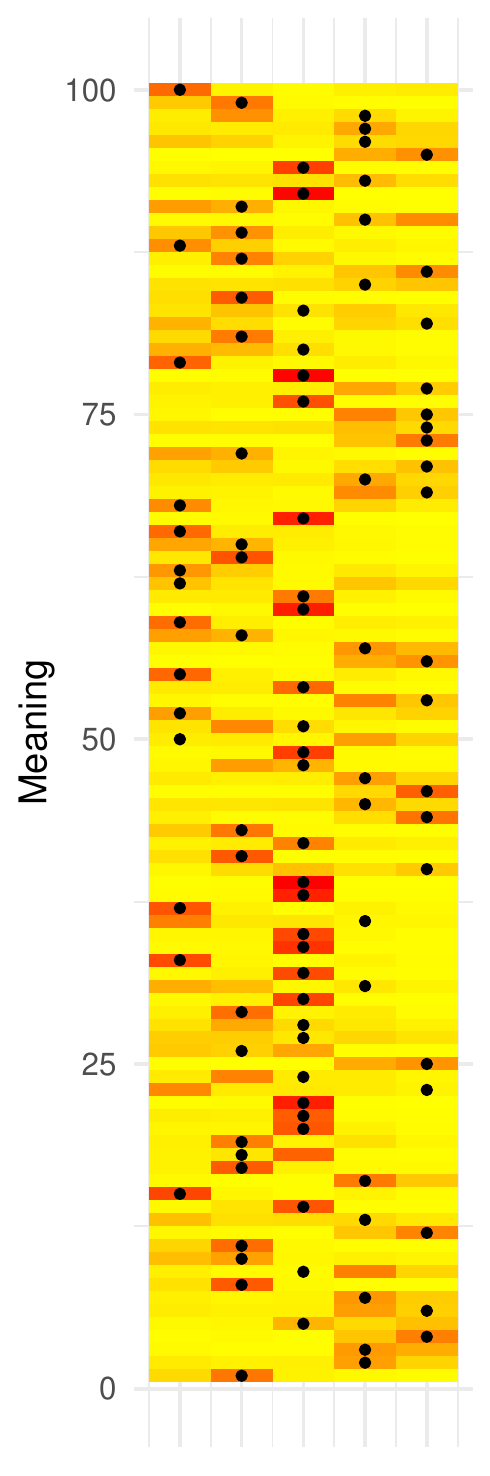} 
\includegraphics[scale=1]{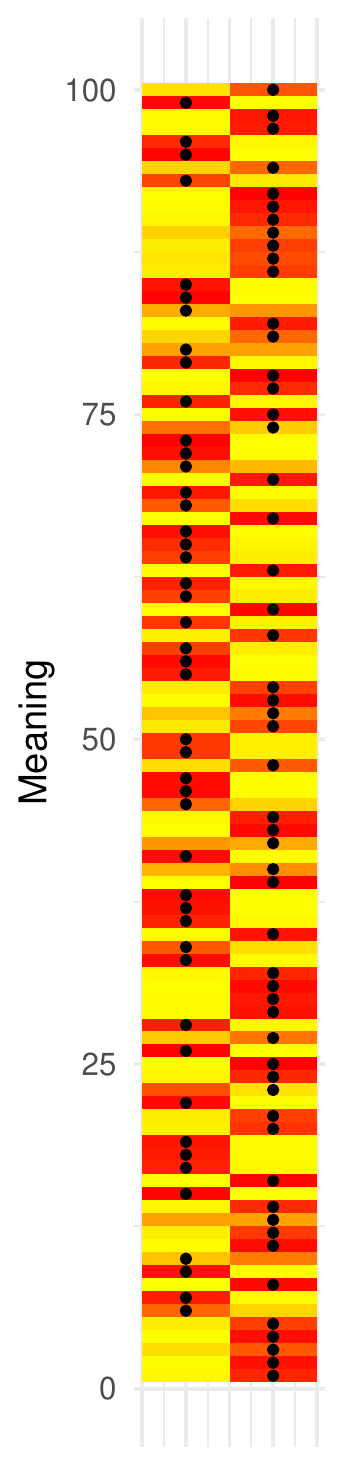} 
\caption{Posterior distribution of internal ancestral values for the first and fourth character at the nearest common ancestor of $6$ and $2$. The color represents the posterior probability for each value, the true value corresponds to the black dot.}
\label{fig:simu_reconstruct}
\end{figure}

Concerning the reconstructed ancestral state, the result highly depends on the age of the common ancestor. Inferring the state of the common ancestor to $9$ and $10$ is fairly simple. More interestingly, we present in Figure \ref{fig:simu_reconstruct} the posterior probability for the values of characters $1$ and $4$ at the common ancestor between $2$ and $6$, this node is correctly reconstructed only with $87\%$. If the majority of the values are correctly reconstructed, some are completely off. Noticeably, the first character, that evolves faster, is more uncertain than the fourth. We advise to use this reconstruction method with caution, as the efficiency of the method can greatly vary between meanings and characters.

\subsubsection{Other prior}

\begin{figure}
\center
\includegraphics[scale=.4]{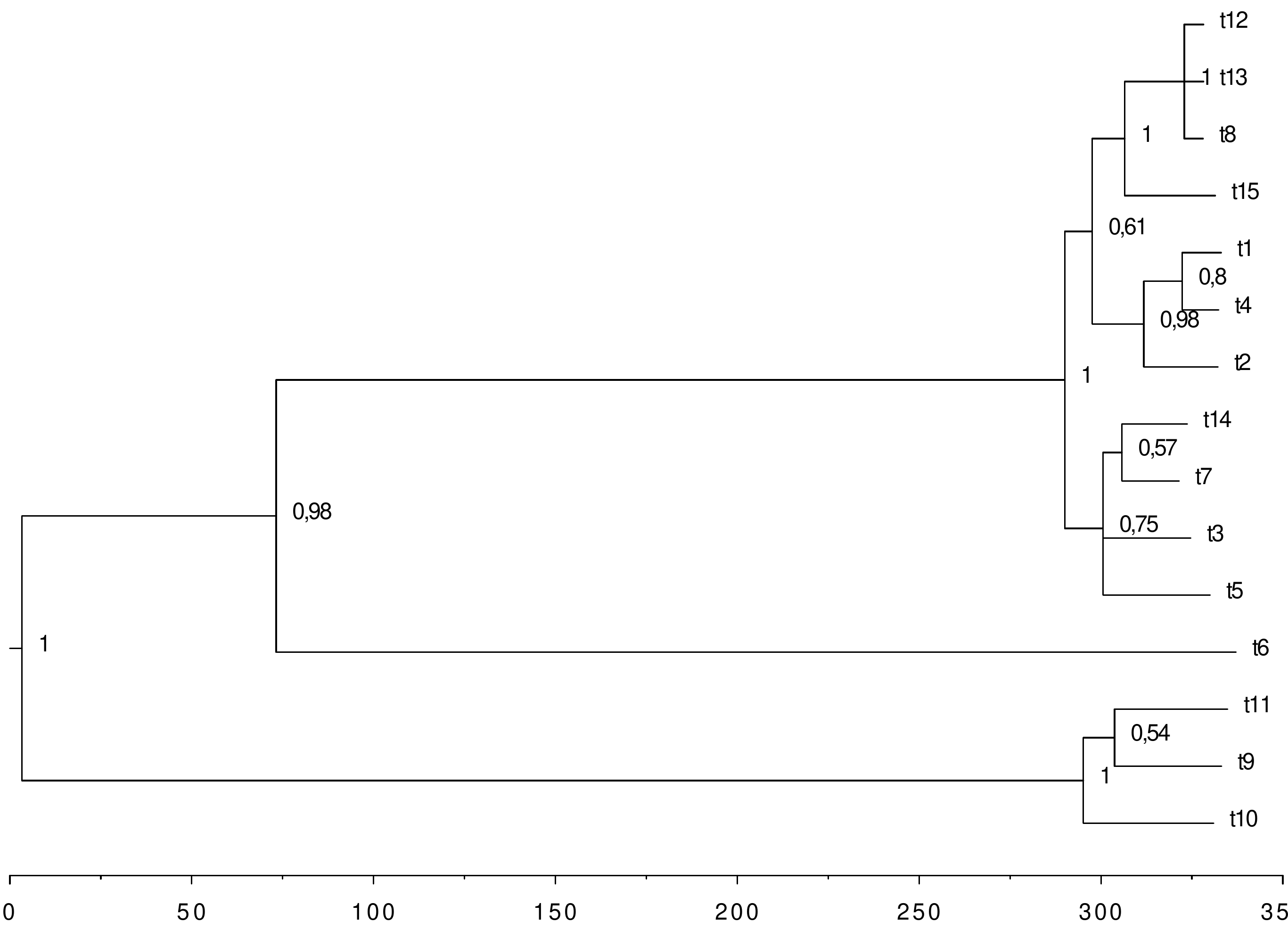}
\includegraphics[scale=1]{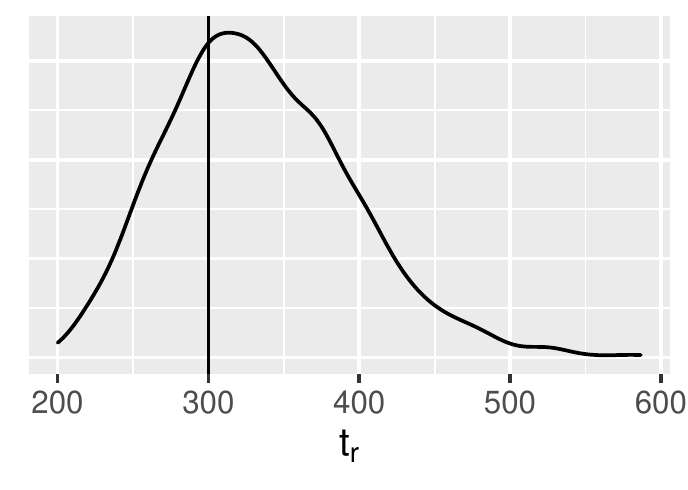}
\caption{Consensus tree and root age posterior distribution for the Gamma prior.}
\label{fig:autrepri}
\end{figure}

We tried a more classical prior on the tree, with the same specifications as previously. Leading to the results presented in Figure \ref{fig:autrepri}

\subsection{Misspecifications of the model}

Our model is clearly an oversimplification of the reality. We propose here to test the resilience of the model to some of the most common

\subsubsection{Correlated traits}


We ran the experiment on the same tree, with the same characters and values --- although with a newly generated dataset, with the particularity that the values for the second character do not come from the above described dynamic. At the end of each edge, the value for the second character is the same as the first one with some noise.

\begin{figure}
\center
\includegraphics[scale=.4]{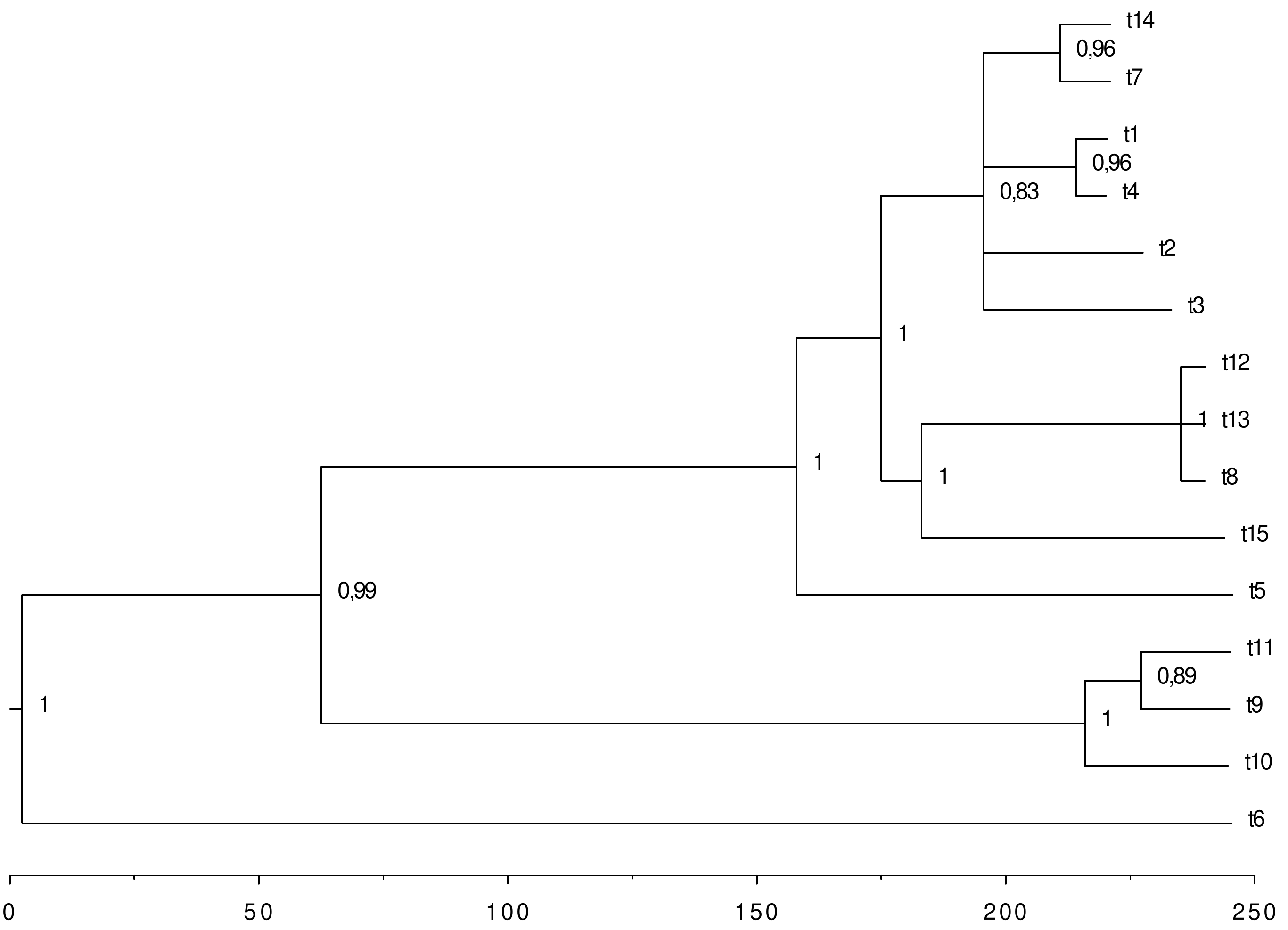} 
\caption{Consensus tree for the correlated dataset test.}
\label{fig:ling:simucorr}
\end{figure}

The resulting consensus trees are presented in Figure \ref{fig:ling:simucorr}. Overall the inference does not seem very much affected, the tree posterior is strikingly concentrated around the consensus topology. The parameters associated with the second character seem a little bit off, which may be strange as \emph{both} characters are correlated.

\begin{figure}
\center
\includegraphics[scale=.9]{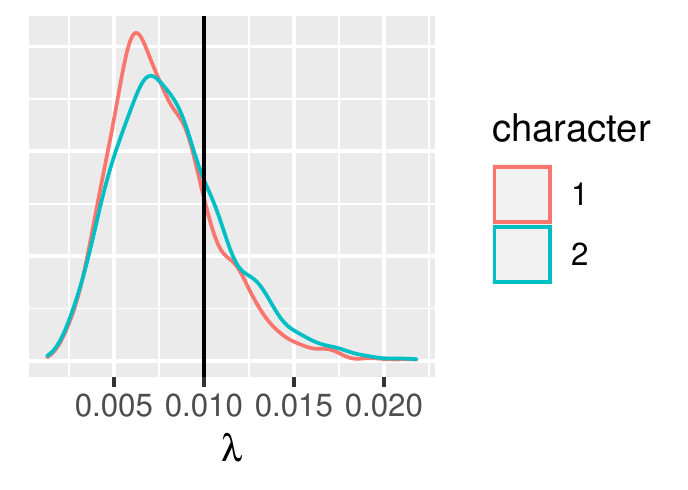} \includegraphics[scale=.9]{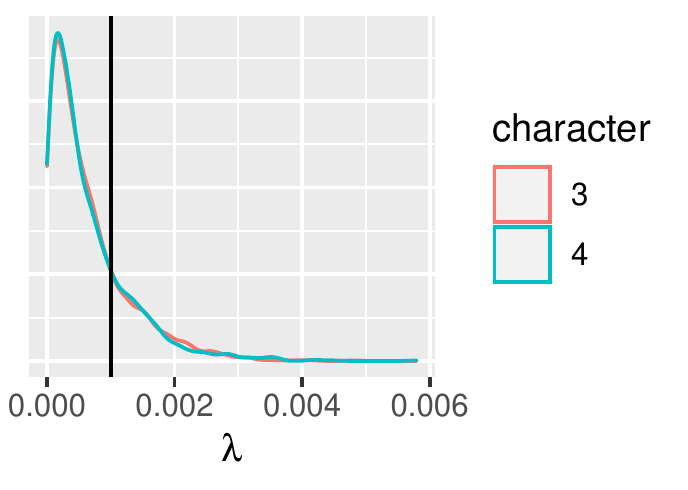}  \includegraphics[scale=.9]{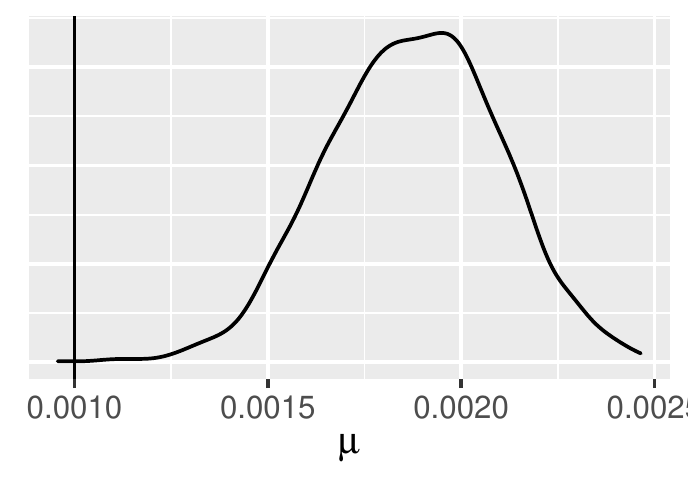}  \includegraphics[scale=.9]{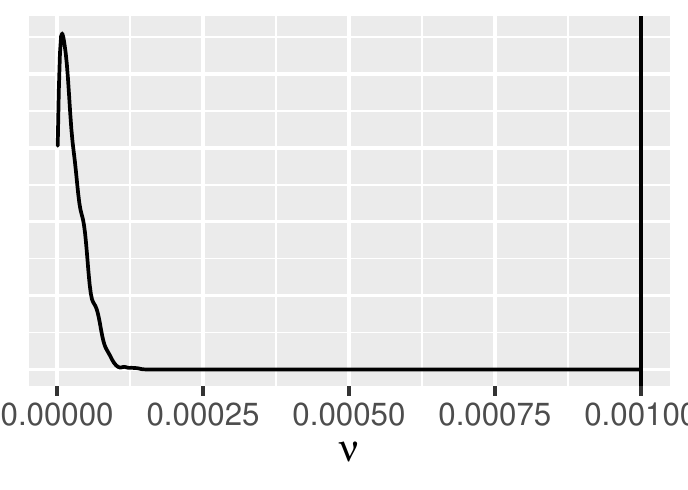} \includegraphics[scale=.9]{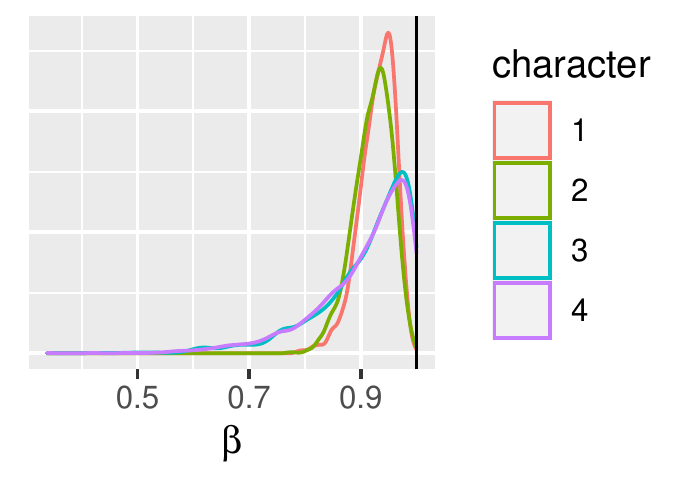} 
\caption{Posterior estimations of $\lambda$, $\rho$, $\nu$, $\beta$ ant the root age. The true values are indicated by the vertical lines.}
\label{fig:simusparamcorr}
\end{figure}

\subsubsection{Selection of transformations}

If the set of possible transformations in the implementation is larger than the true one, we can assume that the inference will lose in efficiency. In the previous model, it is clear that the likelihood of the model is drastically changed if the possible transformations are changed. The identifiability of the model is even unclear, as there are several parameters $p$ that can lead to a same likelihood.

We ran our numerical methods on such a model, the parameter of the experiment are the same as the synthetic exact experiment. The only change is that we did not reduce the number of transitions possible for the two first characters (we have 20 possible transitions instead of the 12 truly present).

\begin{figure}
\center
\includegraphics[scale=.4]{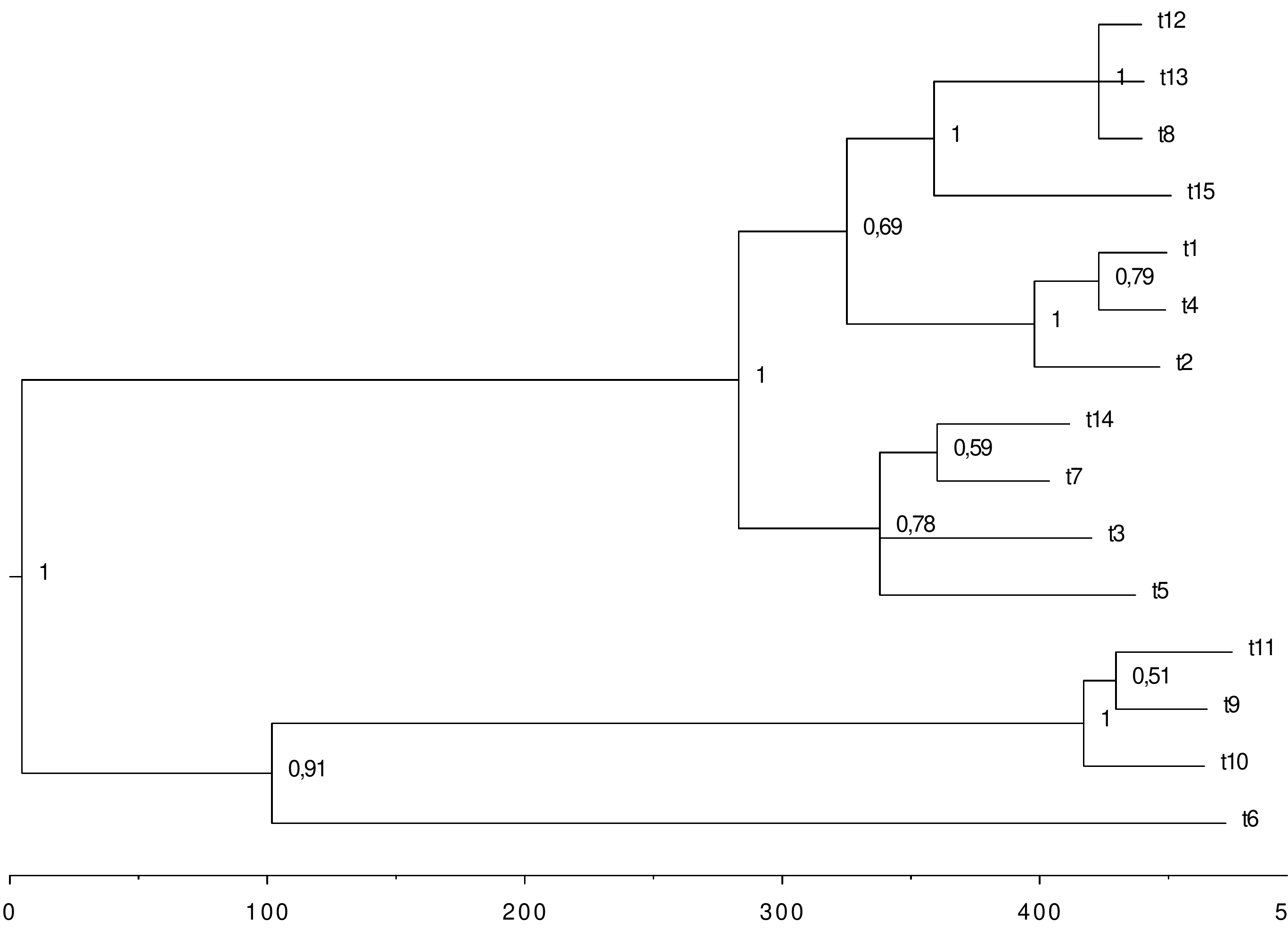} 
\caption{True and consensus tree for the unknown transformations sets}
\label{fig:ling:simu3transf}
\end{figure}

However, this is clearly not efficient if the size of $\mathfrak{R}_k$ is too high. For example, if we allow all the pairwise transformations in a character with $30$ value we get more than a thousand pairs ; the convergence of the numerical methods are then vastly slower and we cannot reach a satisfying equilibrium, and the data for this character can only be explained with noise or apparitions, which is not optimal. We recommend to reduce to less than 30 the size of $\mathfrak{R}_k$.

\begin{figure}
\center
\includegraphics[scale=.9]{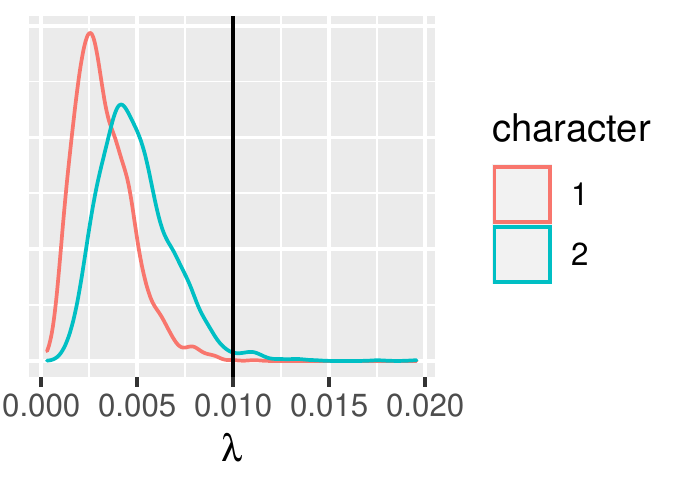} \includegraphics[scale=.9]{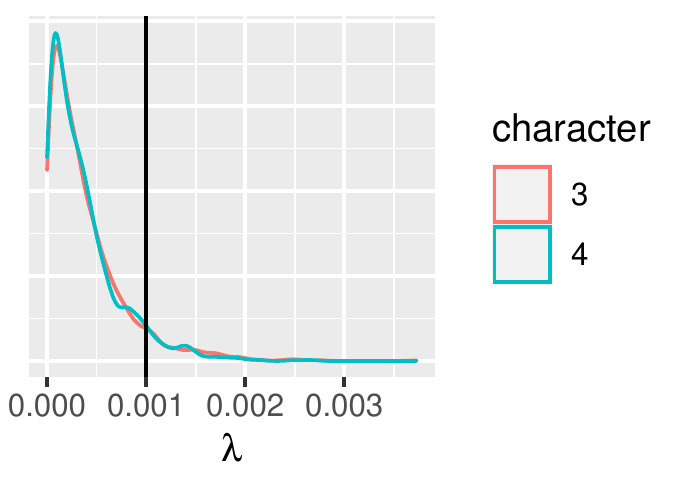}  \includegraphics[scale=.9]{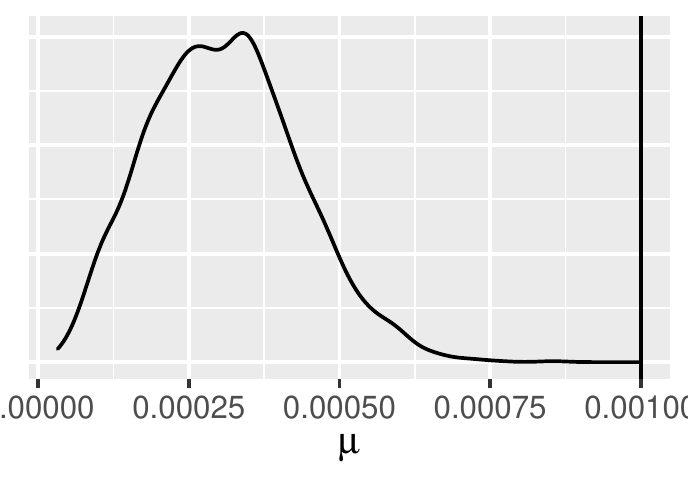}  \includegraphics[scale=.9]{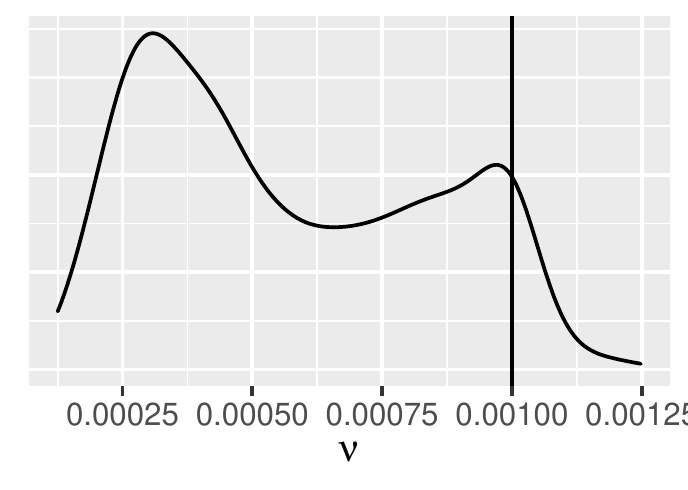} \includegraphics[scale=.9]{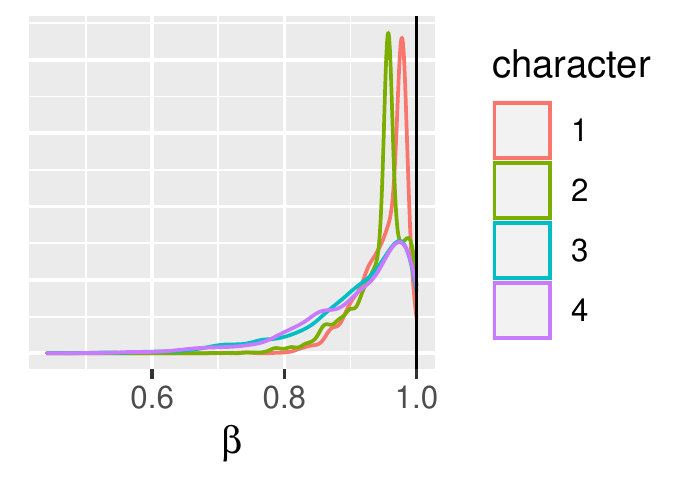}\includegraphics[scale=.9]{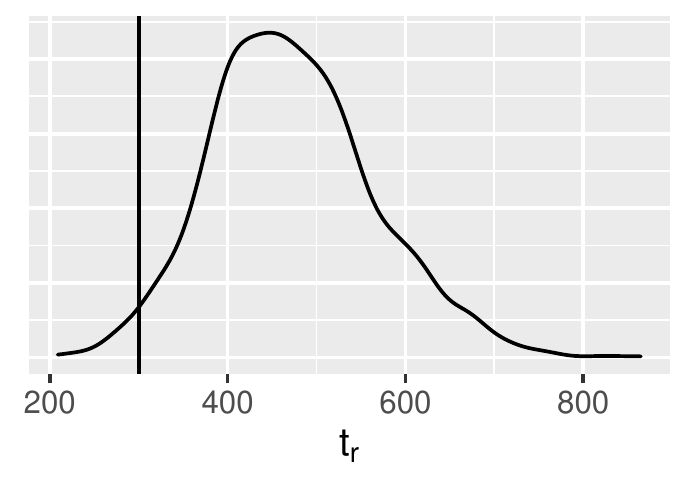} 
\caption{Posterior estimations of $\lambda$, $\rho$, $\nu$, $\beta$ ant the root age for the experiments with unknown $\mathfrak{R}_k$. The true values are indicated by the vertical lines.}
\label{fig:simusparam15transf}
\end{figure}

\begin{figure}
\center
\includegraphics[scale=1]{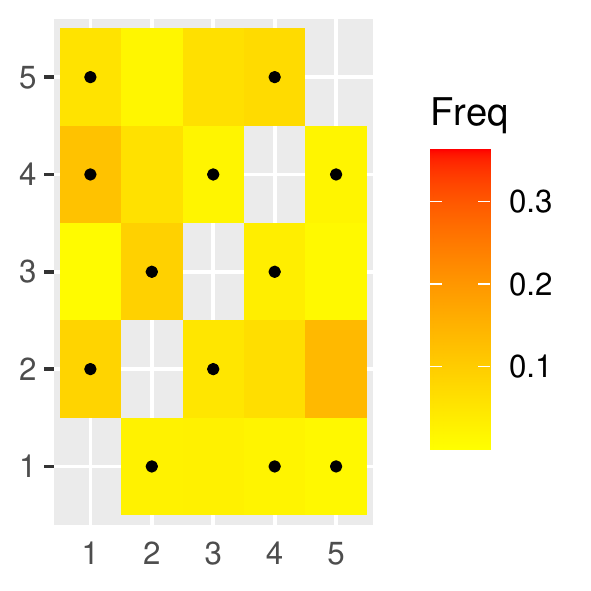} 
\caption{Posterior of the transformation repartition $p$ for the first characters the real transformations are indicated by dots. The color represents the posterior intensity.}
\label{fig:simustransf15transf}
\end{figure}

In these experiments the inference of the other parameter is of lesser quality, with parameter overall underestimated, see Figure \ref{fig:simusparam15transf}. The inference of the probability of the transformations is not quite satisfying, as we can see in Figure \ref{fig:simustransf15transf}, there is no significant difference in the reconstituted probability, this comes certainly from the small number of transformations in total on the tree --- around a dozen.

\subsubsection{Data simulated from a forest}

\begin{figure}
\center
\includegraphics[scale=.4]{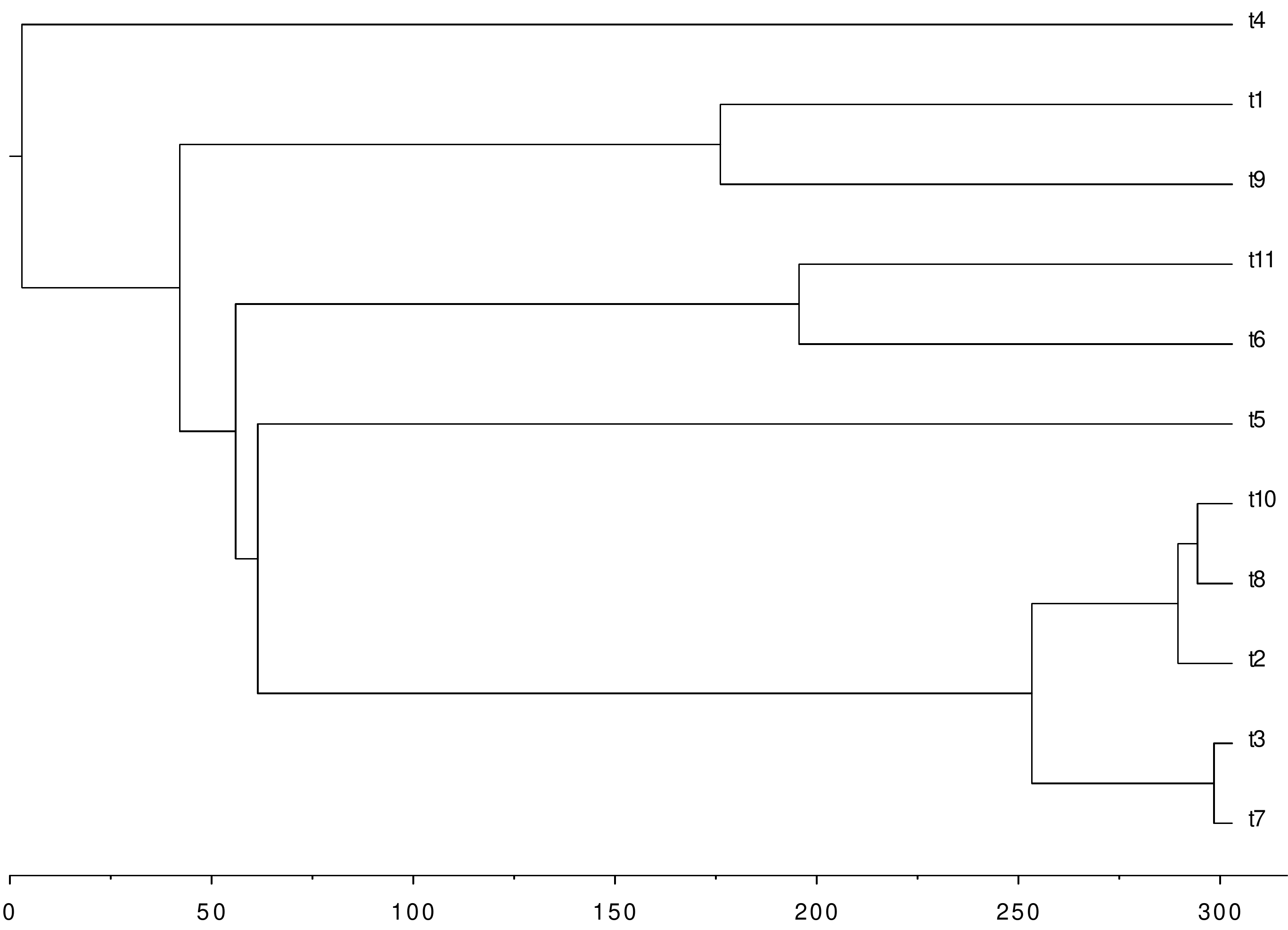}
\includegraphics[scale=.4]{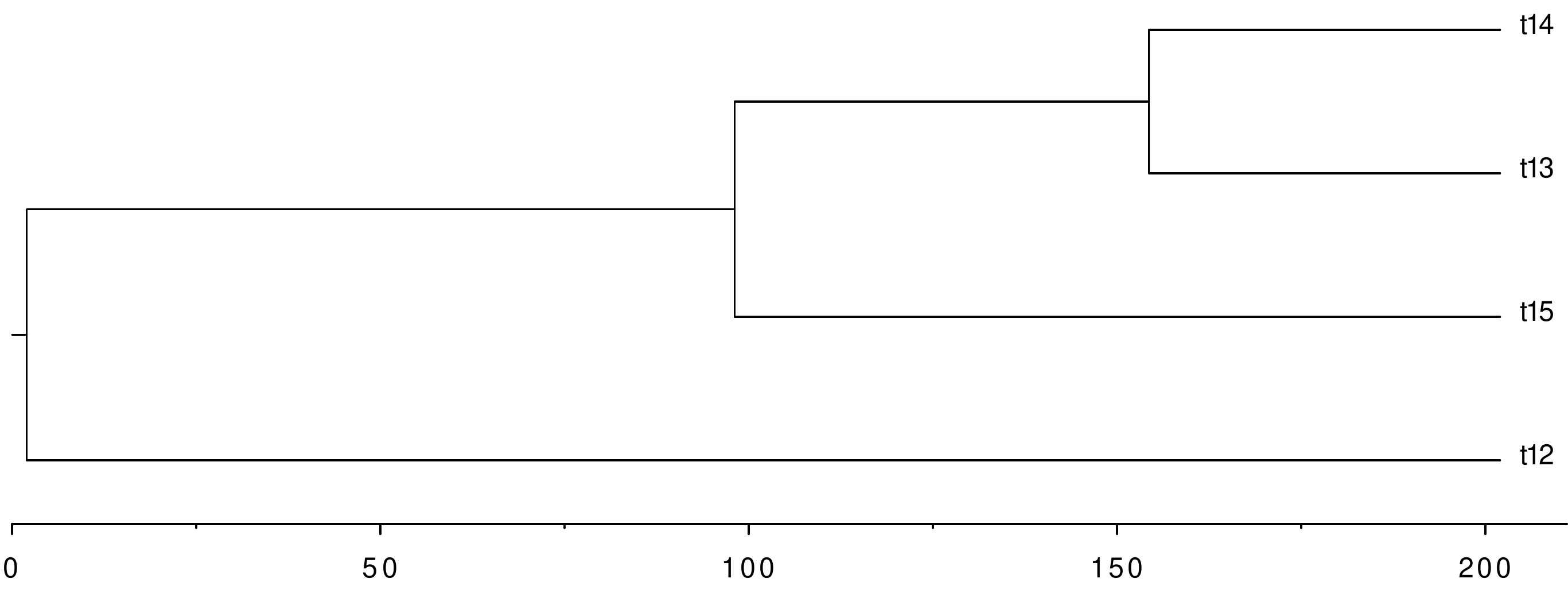} 
\caption{Two trees used for the forest example}
\label{fig:2arbresvrais}
\end{figure}

In this last example, maybe the most interesting for our application, we concatenate the dataset coming from a $11$-leaved tree and a $4$-leaved tree represented in Figure \ref{fig:2arbresvrais}. That is we try to infer a single tree where there should be two. This is a quite important case as we are never sure that some language are indeed related.

\begin{figure}
\center
\includegraphics[scale=.4]{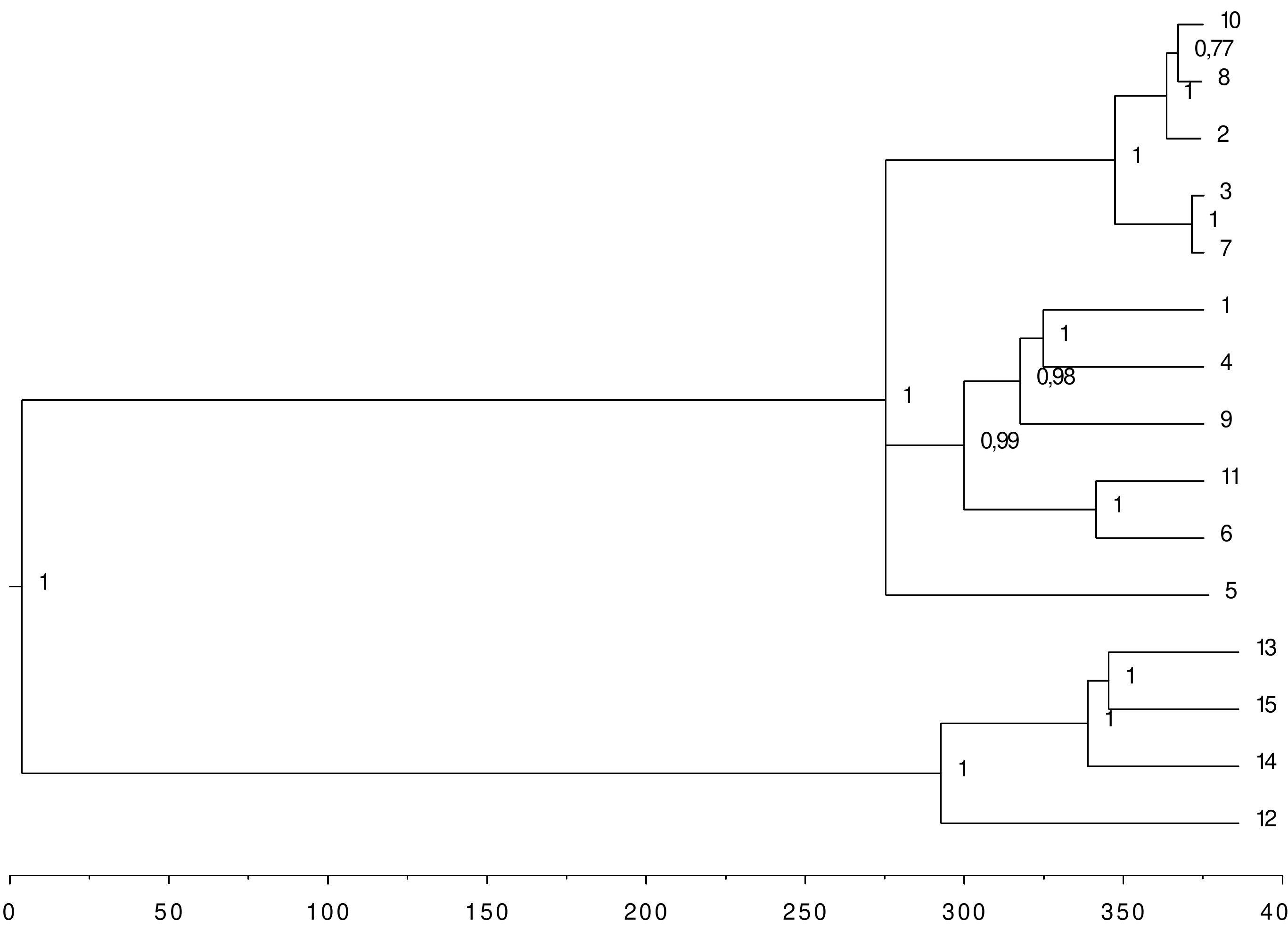}
\includegraphics[scale=1]{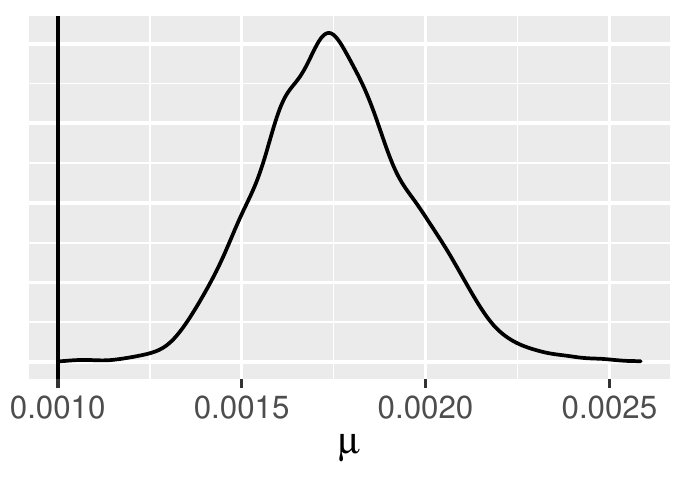} 
\caption{Results of the inference for the topology and the apparition parameter $\rho$ on the forest dataset.}
\label{fig:2arbresvraisres}
\end{figure}

The results are presented in Figure \ref{fig:2arbresvraisres}. Interestingly enough, the two group of language are reconstructed as subtree with rather long edges. Clearly, if the evolution parameter were higher this would not be the case, especially for a small number of characters and a small number of possible value. The evolution parameter however are quite off, as they must be quite important to explain this unique tree. The root is not so old, which is an effect of the prior.

\end{document}